\documentclass[useAMS,usenatbib]{mn2e}
\usepackage{psfig}
\newcommand{\lapp}{\mbox{\raisebox{-0.3em}{$\stackrel{\textstyle <}{\sim}$}}}
\newcommand{\gapp}{\mbox{\raisebox{-0.3em}{$\stackrel{\textstyle >}{\sim}$}}}
\def\kms{km~s$^{-1}$}
\title{Probing radio source environments via H{\sc i} and OH absorption}

\author[Gupta et al.]{Neeraj Gupta$^{1}$\thanks{E-mail: neeraj@ncra.tifr.res.in (NG);
csalter@naic.edu (CJS); djs@ncra.tifr.res.in (DJS); tghosh@naic.edu (TG);
sjk@geofisica.unam.mx (SJ)},
C.J. Salter$^{2}$, D.J. Saikia$^{1,3}$, T. Ghosh$^2$ and S. Jeyakumar$^{4}$ \\
$^{1}$ NCRA, TIFR, Post Bag 3,  Ganeshkhind, Pune 411 007, India \\
$^{2}$ Arecibo Observatory, NAIC,  HC3 Box 53995, Arecibo, Puerto Rico PR 00612, USA \\
$^{3}$ Jodrell Bank Observatory, University of Manchester, Macclesfield, Cheshire SK11 9DL \\
$^{4}$ Instituto de Geofisica, UNAM, Mexico }

\begin{document}

\date{Accepted. Received; in original form }

\pagerange{\pageref{firstpage}--\pageref{lastpage}} \pubyear{2005}

\maketitle

\label{firstpage}

%%%%%%
\begin{abstract}
We present the results of H{\sc i} and OH absorption measurements towards a sample of radio sources
using the Arecibo 305-m telescope and the Giant Metrewave Radio Telescope (GMRT).  In total, 27 radio sources
were searched for associated 21-cm  H{\sc i} absorption. One totally new H{\sc i} absorption system was
detected against the radio galaxy 3C258, while five previously known H{\sc i} absorption systems, and one
galaxy detected in emission, were studied with improved frequency resolution and/or sensitivity.
Our sample included 17 gigahertz peaked spectrum (GPS) and compact steep spectrum (CSS) objects, 4 of which
exhibit H{\sc i} absorption. This detection rate of $\sim$25\% compares with a value of $\sim$40\% by
Vermeulen et al. for similar sources.
We detected neither OH emission nor absorption towards any of the sources that were observed at Arecibo.
We are, however, able to estimate a limit on the abundance ratio of
N(H{\sc i})/N(OH)$\gapp4\times10^{6}$ for 3C258.

We have combined our results with those from other available H{\sc i} searches to compile a heterogeneous 
sample of 96 radio
sources consisting of 27 GPS, 35 CSS, 13 compact flat spectrum (CFS) and 21 large (LRG) sources.
The H{\sc i} absorption detection rate is highest ($\sim$45\%) for the compact GPS sources and least for
the LRG sources.  We find H{\sc i} column density to be anticorrelated with source size, as reported earlier
by Pihlstr\"om et al., a trend which is consistent with the results of optical spectroscopy.
The H{\sc i} column density shows no significant dependence on either redshift or luminosity, which are
themselves strongly correlated. These results suggest that the environments of radio sources on GPS/CSS scales
are similar at different redshifts.  Further, in accordance with the unification scheme, the GPS/CSS galaxies have
an H{\sc i} detection rate of $\sim$40\% which is higher than the detection rate ($\sim$20\%)
towards the GPS/CSS quasars.  Also, the principal (strongest) absorption component detected towards GPS sources
appears blue-shifted in $\sim$65\% of the cases.  This is in agreement with the growing evidence for
jet-cloud interactions playing an important role in determining the ionization and kinematical properties of
the ambient gas.
\end{abstract}


%%%%%%%%%%%%%%%%%%%%%%%%%%%%
%%%%%%%%%%%%%%%%%%%%%%%%%%%%%%

%%%%%%%%%%%%%%
\begin{table*}
\caption{Radio sources observed with the Arecibo telescope and GMRT.}
\begin{center}
\begin{tabular}{|l|l|c|l|l|r|l|l|l|l|l|c|l|}
\hline
Source & Alt. & Opt.  & Redshift &S$_{\rm{5GHz}}$ &$\alpha^5_{1.4}$&Spect. &P$_{\rm{5GHz}}$&Str.   & LAS & LLS   & Ref. & Radio \\
name   & name & id.   &          &           &             &~class  &        ~10$^{25}$&class&                      &       &    & class\\
        &      &       &          &   ~Jy      &             &       & W/Hz      &     & ~~$^{\prime\prime}$ &  kpc  &  &\\
   ~~(1)   & (2)  &  (3)  &  ~~(4)     &   ~(5)     &     (6)     &  ~(7)  & ~(8)       & (9) & (10) &  (11)  & (12)  & (13) \\
\hline
J0031$-$2652 &MRC     &Q    &0.333    & 0.12 & 1.03& STP & 4.4   & U   &$<$1       &$<$4.8   &   1   &CSS  \\
J0034+3025   &B2      &G    &0.1744   & 0.07 & 1.06& STP & 0.7   & SR  &$\sim$1.3  &2.9      &   2   &CSS  \\
J0040$-$2043 &MRC     &G    &0.091    & 0.34 & 0.51& STP & 0.7   & CE  &$<$2.4(15) &$<$4.0(25.5)&0   &CSS  \\
J0119+3210   &4C+31.04&G    &0.060    & 1.58 & 0.47& LFT & 1.3   & T   &0.08       &0.1      &   4   &CSS  \\
J0137+3309   &3C48    &Q    &0.3700   & 5.73 & 0.82& LFT & 245   & T   &1.2        &6.12     &   5   &CSS  \\
J0251+4315   &S4      &Q    &1.311    & 1.43 &$-$0.45&GPS& 449   & T?  &0.0124     &0.11     &   6   &GPS  \\
J0255$-$2153 &MRC     &G    &0.1128   & 0.17 & 0.97& STP & 0.58  & U   &$<$2       &$<$4.0   &   3   &CSS  \\
J0301+3512   &NGC1167 &G    &0.0165   & 0.86 & 0.58& STP & 0.1   & T?  &1.13       &0.3      &   7   &CSS  \\
J0347$-$2900 &MRC     &G    &0.1423   & 0.30 & 0.90& STP & 1.54  & U   &$<$5       &$<$12.5  &   3   &CSS  \\
J0513+0157   &4C+01.13&BL/G &0.0881$^\dag$&0.14&0.40&FLT$^*$& 0.2& SR  &61.2       &97.9     &   8   &LRG  \\
J0645+2121   &3C166   &G    &0.2449   & 1.24 & 0.58& STP & 19.9  & T   &45.0       &171      &   9   &LRG  \\
J0725$-$0054 &PKS     &G    &0.1273   & 1.38 & 0.01& FLT & 4.9   & CJ  &0.03       &0.07     &   10  &CFS  \\
J0805+2409   &3C192   &G    &0.0598   & 1.86 & 0.88& STP & 1.5   & T   &200        &220      &   11  &LRG  \\
J0822+0557   &3C198   &G    &0.08156  & 0.44 & 1.13& STP & 0.7   & D   &198        &297      &   12  &LRG  \\
J0901+2901   &3C213.1 &G    &0.194    & 0.78 & 0.53& STP & 7.4   & T   &5.7(40)    &18.2(128)&   13  &LRG  \\
J1124+1919   &3C258   &G    &0.165    & 0.45 & 0.60& STP & 3.0   & D   &0.10(60)   &0.28(168)&   7,14&CSS  \\
J1148+5924   &NGC3894 &G    &0.01075  & 0.62 &$-$0.31&FLT& 0.02  & T   &0.04       &0.008    &   15  &CFS  \\
J1347+1217   &4C+12.50&Sy/Q &0.12174  & 3.10 & 0.45& LFT & 10.8  & T   &0.09       &0.20     &   16  &CSS  \\
J1407+2827   &Mrk668  &Sy   &0.0766   & 2.41 & 0.87& GPS & 3.1   & D?  &0.008      &0.01     &   17  &GPS  \\
J1516+0015   &MRC     &G    &0.052    & 1.64 & 0.40& FLT$^*$&0.98& CJ  &0.009      &0.01     &   18  &CFS  \\
J1521+0430   &4C+04.51&Q    &1.296    & 1.19 & 1.05& GPS & 1208  & D   &0.137      &1.15     &   19  &GPS  \\
J1604$-$2223 &MRC     &G    &0.141    & 0.39 & 0.67& GPS & 1.9   & D   &0.013      &0.03     &   18  &GPS  \\
J1643+1715   &3C346   &G    &0.162    & 1.39 & 0.78& STP & 9.3   & T   &13.3       &37.2     &   4,13&LRG  \\
J2058+0542   &4C+05.78&G    &1.381    & 0.36 & 0.98& LFT & 402   & U   &$<$0.4     &$<$3.4   &   20  &CSS  \\
J2250+1419   &4C+14.82&Q    &0.23478  & 1.18 & 0.47& STP & 16.8  & D   &0.2        &0.74     &   21  &CSS  \\
J2316+0405   &3C459   &G    &0.2199   & 1.43 & 0.95& LFT & 19.2  & D   &8.0        &28.0     &   22  &LRG  \\
J2325+4346   &OZ438   &G    &0.145    & 0.92 & 0.63& STP & 4.7   & D   &1.6        &4.0      &   7   &CSS  \\
\hline
\end{tabular}
\end{center}
\begin{flushleft}
Col. 1: source name; col. 2: alternative name;
col. 3: optical identification, where BL = BL Lac object, G =
galaxy, Q = quasar and
Sy = Seyfert galaxy; col. 4: redshift; col. 5: total 5-GHz flux density; col. 6: spectral index
between 1.4 and 5 GHz; col. 7:  integrated radio
spectral class, where STP = steep, i.e. $\alpha\ge$0.5, LFT = low-frequency
turnover, GPS = gigahertz peaked and FLT = flat spectrum;
col. 8: 5-GHz luminosity in the rest frame of the source; col. 9: radio structure classification,
where CE =
unresolved component with extended emission, CJ =
core-jet source, D = double-lobed, SR = slightly resolved, T = triple and U =
unresolved (deconvolved size $\lapp$ a fifth of the resolution element);
cols. 10 \& 11: largest projected angular (LAS) and linear (LLS) size in arcsec and kpc respectively, as measured
from the outermost radio peaks, or from Gaussian fits to slightly resolved sources;
col. 12: references for the radio structure and col. 13:  radio class, where
CSS = compact steep-spectrum, GPS = gigahertz peaked-spectrum, CFS = compact flat-spectrum,
and LRG = large radio source; see text (Section \ref{samp_in}) for further details. \\
References for the largest angular sizes -- 0: This paper;  1: Kapahi et al. (1998b);  2: Saikia et al. (2002);
3: Kapahi et al. (1998a); 4: Cotton et al. (1995); 5: Feng et al. (2005); 6: Fey \& Charlot (2000);
7: Sanghera et al. (1995); 8: NRAO VLA Sky Survey (NVSS) at 1.4 GHz; 9: Neff,
Roberts \&  Hutchings (1995);
10: Bondi et al. (1996);
11: Baum et al. (1988);  12: Fomalont \& Bridle (1978); 13: Akujor \&
Garrington (1995); 14: Strom et al. (1990);
15: Peck \& Taylor (1998);  16: Lister et al. (2003);  17: Stanghellini et al. (2001);
18: Beasley et al. (2002; VCS observations);  19: Xiang et al. (2002);  20: de Breuck et al. (2000);
21: Spencer et al. (1989);  22: Thomasson, Saikia \& Muxlow  (2003) \\
$^{\dag}$ The source was observed at $z = 0.084$, the best available redshift at that time. \\
$^{*}$ The intergrated spectrum exhibits low-frequency steepening. \\
\end{flushleft}
\label{sample}
\end{table*}

%%%%%%%%%%%%%%%%%%%
\section{Introduction}
There is general agreement that the energy output from an active galactic nucleus (AGN) is fuelled by the
supply of gas to the central engine, which is presumably a supermassive black hole (e.g. Rees 1984).
This infall could be triggered by interactions or mergers with companion galaxies, leading to the formation
of circumnuclear starbursts and the fuelling of nuclear black holes (e.g. Sanders et al. 1988).
Although there have been a number of studies of the gas kinematics of nearby AGN in an effort to understand
both the fuelling of this activity and the relationship between the starburst and the AGN, relatively little
is known regarding the situation in powerful radio galaxies. Observationally, there is evidence that the
central regions of such galaxies contain dust plus ionised, atomic and molecular gas which could be reservoirs of
fuel for the central engine.  For example, Walsh et al.  (1989) have shown that long-wavelength infrared emission,
indicative of dust, is correlated with nuclear radio emission. Further, Verdoes Kleijn et al. (1999), from an HST
study of 19 Fanaroff-Riley class I (FRI) galaxies, have demonstrated the presence of dust structures in 17 of these.
Besides such structures, atomic and molecular gas within $\sim$5--10\,kpc of the nuclear region has been
seen in many nearby active galaxies, including several well-known radio galaxies (e.g. Rupen 1997; Lim et al. 2000;
Leon et al. 2003).

%%%%%%%%%%%%%%%%
\begin{table*}
\caption{Observational details and results of the Arecibo search for associated H{\sc i} absorption.
}
\begin{center}
\begin{tabular}{|l|c|c|l|l|l|c|c|}
\hline
Source & S$_{\rm{1.4GHz}}$& Date &  $\Delta v$$^{\rm{H{\sc I}}}$ & $\sigma_{\rm{FA}}^{\rm{H{\sc I}}}$ &  
FA$_{\rm{max}}$ & W$_{\rm{50\%}}$ & N(H{\sc i})\\
name & Jy & &  km s$^{-1}$ & 10$^{-3}$ &  &km s$^{-1}$ &10$^{20}$\,cm$^{-2}$\\
~~(1)   & (2)  &   (3)  & ~(4)   & ~(5)   &  ~(6)   &   (7)   &   (8)   \\
\hline
J0034+3025 & 0.26 & 2002 Dec  & 16.0      & 2.0       & --           & --        & $<$1.16        \\
J0119+3210 & 2.64 & 2002 Dec  & 0.72$^{a}$& 1.5$^{a}$ &  0.038$^{b}$ & 153$^{b}$ & ~~12.2$\pm$0.14$^{c}$   \\
            &      &          &           &           &  0.030$^{d}$ & 7.6$^{d}$ &                \\
J0301+3512 & 1.84 & 2002 Dec  & 24.0      & 1.3       & --           & --        & $<$0.75        \\
J0513+0157 & 0.35 & 2004 Mar  & 13.6      & 2.7       & --           & --        & $<$1.56        \\
J0645+2121 & 2.59 & 2002 Dec  & 18.0      & 0.85      & --           & --        & $<$0.49        \\
            &      & 2004 Sep &           &           &              &           &                \\
J0725$-$0054&1.40 & 2002 Dec  & 14.8      & 3.6       & --           & --        & $<$2.08        \\
J0805+2409 & 5.33 & 2002 Dec  & 13.0      & 0.62      & --           & --        & $<$0.36        \\
J0822+0557 & 1.98 & 2002 Dec  & 13.6      & 1.4       & --           & --        & $<$0.81        \\
J0901+2901 & 2.00 & 2002 Nov  & 16.5      & 0.46      & --           & --        & $<$0.27        \\
            &      & 2002 Dec &           &           &              &           &                \\
J1124+1919 & 0.88 & 2002 Nov  & 3.5       & 1.3       & 0.095        & 31.5      & ~~5.05$\pm$0.13\\
            &      & 2003 Jan &           &           &              &           &                \\
J1347+1217 & 5.40 & 2002 Nov  & 3.2       & 0.87      & 0.012        & 135       & ~~3.03$\pm$0.16\\
            &      & 2003 Jan &           &           &              &           &                \\
J1407+2827 & 0.82 & 2002 Nov  & 13.4      & 0.66      & 0.003        & 326       & ~~1.53$\pm$0.29\\
            &      & 2003 Jan &           &           &              &           &                \\
J2250+1419 & 1.97 & 2002 Dec  & 17.7      & 0.40      & --           & --        & $<$0.23        \\
J2316+0405 & 4.68 & 2002 Dec  & 17.2      & 0.49      & 0.0034       & 166       & ~~1.33$\pm$0.27\\
            &      & 2005 Aug &           &           &              &           &                \\
\hline
\end{tabular}
\end{center}
\begin{flushleft}
Col. 1: source name; col. 2: 1.4-GHz flux density from NASA/IPAC
Extragalactic Database (NED); col. 3: dates of observation;
col. 4 \& 5: the smoothed velocity resolution and the corresponding 
1-$\sigma$ noise for the H{\sc i} fractional absorption;
cols. 6 \& 7: peak optical depth and full width at half maximum (FWHM) for the absorption components
derived directly from spectra with the velocity resolution given in col.
5, and col. 8: H{\sc i} column density assuming T${\rm _s}$=100 K.  Upper limits
and quoted errors are both 3$\sigma$.  The upper limits assume
$\Delta v$=100 km~s$^{-1}$ and $f_c$=1.0. \\
\hspace*{0mm} $^{a}$ These values are for a spectrum with a total
bandwidth of 3.125 MHz. The values for a total bandwidth of 25 MHz
were $\sigma_{\rm{FA}}^{\rm{H{\sc I}}}$= 0.00051 and $\Delta v$$^{\rm{H{\sc I}}}$=5.8 km s$^{-1}$. \\
\hspace*{0mm} $^{b}$ The maximium optical depth and velocity half
width given represent just the main absorption component. \\
\hspace*{0mm} $^{c}$ The integrated column density is for the
total absorption spectrum. \\
\hspace*{0mm} $^{d}$ The maximium optical depth and velocity half
width given represent just the secondary absorption component.
\end{flushleft}
\label{obsArres}
\end{table*}
%%%%%%%%%%%%%%%%%%%%%%%%%%

In the canonical model of an AGN, the nuclear region contains a torus around a supermassive black hole.
The torus can consist of ionised, atomic and molecular gas components. An understanding
of the distribution and kinematics of the different components of the circumnuclear gas is important
both for studying the anisotropy in the radiation field and thereby testing the unified scheme, and for
understanding the fuelling of the radio activity.  In addition, detection of this gas over a large range of
redshifts and source sizes can provide valuable information on the evolution of its properties
with redshift and  source size (age).  At radio wavelengths, the ionized component of this gas may be
probed via radio polarization measurements of compact source components residing within the
dense interstellar environments of their parent galaxies. Radio cores, and compact steep-spectrum (CSS)
and gigahertz peaked-spectrum (GPS) sources constitute such objects. CSS sources are defined
as those with a projected linear size $\lapp$15 kpc (H$_o$=71 km s$^{-1}$ Mpc$^{-1}$, $\Omega_m$=0.27,
$\Omega_\Lambda$=0.73, Spergel et al. 2003) and having a steep high-frequency radio spectrum
($\alpha\gapp$0.5, where S($\nu)\propto\nu^{-\alpha}$).  The structural and polarization asymmetries
observed in these sources require an asymmetric distribution of gas in the central regions of the galaxy,
which may be related to the infall of gas that fuels the radio source (Saikia et al. 1995;
Saikia \& Gupta 2003, and references therein).

An important way of probing atomic gas on sub-galactic scales is via H{\sc i} absorption towards the compact
components of CSS and GPS sources or the radio nuclei of larger objects (e.g. van Gorkom et al. 1989;
Conway \& Blanco 1995; Peck et al. 2000; Pihlstr\"om 2001; Vermeulen et al. 2003; Pihlstr\"om, Conway \&
Vermeulen 2003).  A pioneering study of H{\sc i} absorption for a well-defined sample of radio galaxies
was made by van Gorkom et al. (1989). This detected H{\sc i}  in 4 out of 29 galaxies, all 4 being dominated
by a compact nuclear radio source. More recently, a large survey of CSS and GPS sources detected
H{\sc i} absorption in 33\%, with the H{\sc i} column density being anti-correlated with source size (Vermeulen et
al.  2003; Pihlstr\"om et al. 2003).  The H{\sc i} spectra exhibit a variety of line profiles, implying significant,
sometimes complex, gas motions.  van Gorkom et al. (1989) had reported that the H{\sc i}-absorption features tend
to be  redshifted from the systemic velocity, suggesting infall of gas. However, recent observations show a more
complex situation. For example, Vermeulen et al. find many sources with substantial red and blue shifts,
suggesting that atomic gas may be flowing out or falling in, interacting with the jets, or rotating
around the nucleus.

Although many recent H{\sc i} absorption studies have been based on samples of CSS and GPS objects,
Morganti et al. (2001) observed extended radio galaxies from the 2-Jy sample of Wall \& Peacock (1985),
finding the following trends. H{\sc i}  absorption was detected in only 1 of the 10 FRI radio galaxies observed.
For FRII radio galaxies, they detected H{\sc i}  absorption in 3 of 4 NLRGS, while absorption was not detected
in any of 4 BLRGs.  Although this is largely consistent with the predictions of the unified scheme
(see van Ojik et al. 1997; Pihlstr\"{o}m et al. 2003), the H{\sc i}  is blueshifted from the systemic
velocity in 2 of the 3 NLRGs, suggesting that attributing the absorption to just a torus may be too simplistic.

\begin{table*}
\caption{Observational details and results of the GMRT search for associated H{\sc i} absorption.}
\begin{center}
\vspace{5mm}
\begin{tabular}{|l|l|r|c|c|r|c|c|c|c|c|c|}
\hline
Source&~~Date&Obs. freq& BW &t$_{\rm{int}}$&  Beam~~~   &   Peak flux & $\sigma$ & $\sigma_{\rm{FA}}^{\rm{H{\sc I}}}$ & 
FA$_{\rm{max}}$ & W$_{\rm{50\%}}$ & N(H{\sc i})\\
name  &    &  MHz~~   & MHz &   hr &$^{\prime\prime}\times^{\prime\prime}$ $^\circ$~~~ &  Jy/b  &  mJy/b/ch  & 10$^{-3}$ &  & km s$^{-1}$ & 10$^{20}$cm$^{-2}$\\
~~(1)   & ~~~(2)  &  (3)~~~   & (4)  &  (5)  &   (6)~~~~~    &  (7)   &  (8)   &   (9)   &    (10) & (11) & (12)  \\
\hline
J0031$-$2652 & 2003 Dec & 1065.570& 4 & 3.5 & 3.9$\times$2.8 ~19 & 0.55  & 1.8     &  $<$3.27 & --  & --     & $<$1.90 \\
J0040$-$2043 & 2003 Dec & 1301.930& 4 & 2.5 & 4.2$\times$3.0 ~~3 & 0.40  & 0.6     & $<$1.50  & --  & --     & $<$0.87 \\
J0137+3309$^\ast$ &  ~~~-- &  --     & 4 &  -- &  --                & 20.19~ & 20 & $<$1.00  & --  & --     & $<$0.06 \\
J0251+4315 & 2003 Jul & 614.894 & 2 & 1.0 & 13.2$\times$6.9 ~~4 & 0.80  & 3.6      & $<$4.50  & --  & --   & $<$2.61 \\
J0255$-$2153 & 2003 Dec & 1276.425& 4 & 3.0 & 3.1$\times$2.5 ~30 & 0.44  & 2.8     & $<$6.36  & --  & --   & $<$3.69 \\
J0347$-$2900 & 2003 Dec & 1243.461& 4 & 3.0 & 6.6$\times$2.5 ~43 & 2.21  & 1.6     & $<$0.72  & --  & --   & $<$0.42 \\
J1148+5924 & 2003 Jun & 1405.299& 4 & 1.5 & 4.9$\times$2.8   ~34 & 0.62  & 0.6     & $<$0.97  & 0.034&157 & ~~9.66$\pm$3.24 \\
J1516+0015 & 2003 Dec & 1350.196& 4 & 3.5 & 5.1$\times$4.9 ~59 & 0.68  & 0.6       & $<$0.82  & --  & --   & $<$0.51 \\
J1521+0430 & 2003 Jul & 618.644 & 2 & 1.3 & 29.3$\times$8.0  43 & 4.34  & 4.8      &  $<$1.11 & --  & --      & $<$0.64 \\
  & &         &   & &                          & 0.41 & &         & & & \\
J1604$-$2223 & 2003 Jul & 1244.878& 4 & 2.3 & 6.4$\times$3.1   ~35 & 0.72  & 4.1     & $<$5.69& --  & --       & $<$3.30 \\
J2058+0542 & 2003 Jul & 596.558 & 2 & 2.0 & 9.5$\times$6.2   172 & 2.04  & 5.0     &  $<$2.45 & --  & --    & $<$1.42 \\
J2325+4346 & 2003 Jan & 1240.529& 8 & 3.2 & 3.0$\times$1.9   ~56 & 0.48  & 0.7     & $<$1.45  & --  & --     & $<$0.84 \\
\hline
\end{tabular}
\end{center}
\begin{flushleft}
Col.  1: source name; col. 2:
year and month of observations; col. 3:  redshifted 21-cm frequency;
col. 4:  baseband bandwidth (BW);
col.  5:  observing time in hr (excluding
calibration overheads); cols. 6 \& 7: restoring beam and peak
brightness in Jy/beam for the continuum image made using
line-free channels; col. 8: rms noise in the spectrum in units of mJy/beam/channel; 
col. 9: 1-$\sigma$ noise for the H{\sc i} fractional absorption;
cols. 10 \& 11: peak optical depth and FWHM for the absorption components
derived directly from the spectra,  
and col. 12:  H{\sc i} column density in units of
10$^{20}$ cm$^{-2}$, assuming T${\rm _s}$=100 K; upper limits are 3$\sigma$ values,
and assume $\Delta v$=100 \kms and $f_c$=1.0.  The values of noise and peak optical
depth have been estimated from the unsmoothed spectra except for J1148+5924 where
the values have been estimated from the spectrum shown in Fig.~\ref{specngc3894}. \\
$^{\ast}$ See text and Table ~\ref{GMRTobs3C48} for details of the observations of J0137+3309 (3C48).
\end{flushleft}
\label{GMRTobs+res}
\end{table*}
% %%%%%%% %% %%%%%%%%%
%% %% %

%%%%%%%%%%%%%%%%%%%%%%
In order to extend these H{\sc i} investigations to a larger number of GPS and CSS objects, especially those
that are of lower luminosity or more distant, and also towards larger sources, we have begun a program using
the Arecibo 305-m telescope and the Giant Metrewave Radio Telescope (GMRT).  In addition to H{\sc i},
at Arecibo we have attempted to detect the OH radical which has so far received little attention in
this context. In this paper we present the results of the first phase of our study and examine the dependence
of H{\sc i} column density on projected linear size, redshift, luminosity and host galaxy type by combining
our results with those available in the literature. The observations are described in Section
2, while the results, with notes on individual sources, are presented in Section 3.
In Section 4, we combine our results with those of similar H{\sc i} searches and discuss the statistical
trends in gas properties with respect to radio-source characteristics and redshift.
The results are summarized in Section 5.

%%%%%%%%%%%%%%%%%%%%%%%%%%%%%%%%%%%%%%%%%%%%%%%%%%%%%%%%%%%%
\section{Observations}

%%%%%%
\subsection{The observed sample}
\label{samp_in}
The principal targets of the Arecibo and GMRT observations were CSS and GPS sources 
whose redshifted 21-cm H{\sc i} lines lie in the observing bands of these telescopes.  
In total, we observed 17 GPS and
CSS sources with redshifts of $z\lapp$1.4. We also observed 7 large radio galaxies (LRGs; sizes $\gapp$15 kpc),
and 3 compact flat-spectrum (CFS) objects ($\alpha\lapp$0.5 and sizes $\lapp$15 kpc). At Arecibo, data were
also recorded for the OH main-lines in an attempt to detect these in either emission or absorption.
The sources and some of their properties are listed in Table~\ref{sample}. 
They form a hetergeneous set having been selected from a number of 
samples rather than a single well-defined complete sample. Possible selection effects arising 
from this are discussed later. 

%%%%%
\subsection{Arecibo observations and data reduction}
The 15 radio sources observed at Arecibo consist of both GPS and CSS
objects, and large FRI and FRII sources.  The 7 CSS and GPS sources
that formed the basic Arecibo sample were selected from well-defined
samples of such objects (Fanti et al. 1990; Sanghera et al. 1995, and
references therein) and a sample of compact symmetric objects (Taylor,
Readhead \& Pearson 1996).  They have angular sizes of
$\lapp$1$^{\prime\prime}$, with L-band flux densities in the range
0.3$-$5.4 Jy, and $z\le$0.3.  Seven LRGs and one CFS object, all within
the same flux density and redshift ranges, were added to fill gaps in
the observing schedule. Six of the LRGs were chosen from the 3C
sample.  In addition to the H{\sc i} observations, we made use of the
versatility of the Arecibo spectrometer to simultaneously search for OH
in all objects. Sadly, a number had their OH spectra corrupted by radio
frequency interference (RFI).  See Table~\ref{obsArres} for the
observing log.

%%%%%%
The Arecibo feed platform and its support cables cause blockage to the
incoming wave-front that is focussed on the telescope feed, and also
scatter significant amounts of radiation from directions outside of the
telescope main beam.  The standing-wave pattern resulting from
multipath scattering adds baseline ripples to a line spectrum.  For
standard Arecibo spectral-line observations of sources emitting little
continuum radiation, such effects are minimized via a variant of the
position-switching technique.  However, when the target emits
significant continuum radiation, the standing-wave pattern due to the
continuum emission from the target itself is not cancelled by
subtracting the position-switched OFF from the ON, and a standing-wave
residual remains whose amplitude is proportional to the source
intensity.  Briggs, Sorar \& Taramopoulos (1993) extended the basic
position switching method such as to minimize the effect of these
residual standing waves.  To achieve this, another strong continuum
source (of different redshift) is also observed via position
switching.  Ghosh \& Salter (2002) explored this ``Double
Position-Switching'' (DPS) approach at Arecibo for sources having
significant continuum emission and found it to be very effective in
providing flat spectral baselines.  We therefore adopted DPS for the
present Arecibo observations of our sample, all of which emit
significant continuum radiation.

The present observations recorded the orthogonal linear
polarizations of both the $\lambda$21-cm H{\sc i} line, and the
$\lambda$18-cm OH main-lines.  For all sources except J0119+3210 (see
below), a bandwidth of 12.5 MHz was processed into 2048 channels, with
data acquisition using 9-level quantization.  The typical flux-density
ratio for the DPS source pairs was 2:1.  Weighted averages of the
data for individual DPS cycles yielded the final spectra, which were
Hanning smoothed.  H{\sc i} absorption was detected in five of the
targets (J0119+3210, J1124+1919, J1347+1217, J1407+2827 and
J2316+0405), and emission in a sixth (J0301+3512).  Neither OH emission
nor absorption was detected for any source.  The rms noises on the
H{\sc i} and OH spectra of fractional absorption for a given (often
smoothed) velocity resolution are listed in Tables~\ref{obsArres} and
\ref{obsArOH}.  The peak fractional absorptions, integrated column
densities and full-width half maxima for the five sources with detected
H{\sc i} absorption are given in Table~\ref{obsArres}.

%%%%%%%%%%%%%%%%%%%%%%%%%%%%%%%%%%%%%%%%%%%%%%%
\subsection{GMRT observations and data reduction}
Of the 12 radio sources observed with GMRT to search for associated H{\sc i} absorption, 9 were selected
from the sample of  Snellen et al. (2002) and the Molonglo Reference Catalogue (MRC; Large et al. 1981;
Kapahi et al. 1998a,b). Of the remaining three, two are the CSS objects J0137+3309 (3C48) and J2325+4346,
and the third is the flat-spectrum source J1148+5924 (NGC3894).   The angular size of each source is less than the
spatial resolution of the observations.
%%

%%%%%%
\begin{table}
\caption{Log for the GMRT observations of J0137+3309 (3C48).}
\begin{center}
\begin{tabular}{lccc}
\hline
Obs. run  & Date         & ~t$_{int}$$^*$ & Obs. freq.  \\
&              &  hr          &  MHz       \\
\hline
1        & 2003 Oct 11    & 3.0 & 1039.8 \\
2        & 2003 Dec 10    & 5.0 & 1040.6 \\
3        & 2004 Aug 24    & 7.0 & 1037.7 \\
4        & 2004 Aug 25    & 6.0 & 1037.7 \\
\hline
\end{tabular}
\end{center}
* Integration time includes calibration overheads.  \\
\label{GMRTobs3C48}
\end{table}
H{\sc i} absorption profiles for CSS sources exhibit a variety of shapes and can be as broad as several 100 \kms.
Further, the profiles are often either blue- or red-shifted by a few 100 \kms ~with respect to the systemic velocity
derived from the optical emission lines.  To account for these factors, we used a baseband bandwidth (BB BW) of
8 MHz in our first observing run.  However, we soon realised that such broad bandwidths are both more difficult
to calibrate, and susceptible to RFI.  Thus for subsequent runs we used a BB BW of 4 MHz for observations at
L-band, which has linearly-polarized feeds,
and 2 MHz for those in the 610-MHz band, which has circularly-polarized feeds.  This also gave the option of
using the 6-MHz IF filter which helped
minimise the effects of RFI.  The GMRT FX correlator splits the BB BW into 128 frequency channels.
The above choice of BB BWs thus implies a velocity resolution of $\sim$7 \kms, with a total typical velocity
coverage of $\sim$1000 \kms, narrower than that of Vermeulen et al. (2003) by a factor of $\sim$2.
Details of the GMRT observations are given in Table~\ref{GMRTobs+res}.  GMRT calibration requires observations of
standard flux density and phase calibrators.  In all our observations, except for 3C48, the phase calibrator was
observed approximately every 30 min.  For bandpass calibration, a flux density calibrator (3C48,
3C147 or 3C286) was observed every 2$-$3 hr.

%%%%%%%%%%%%%%%%%%%%%%%
%
\begin{figure*}
\centerline{\vbox{
\psfig{figure=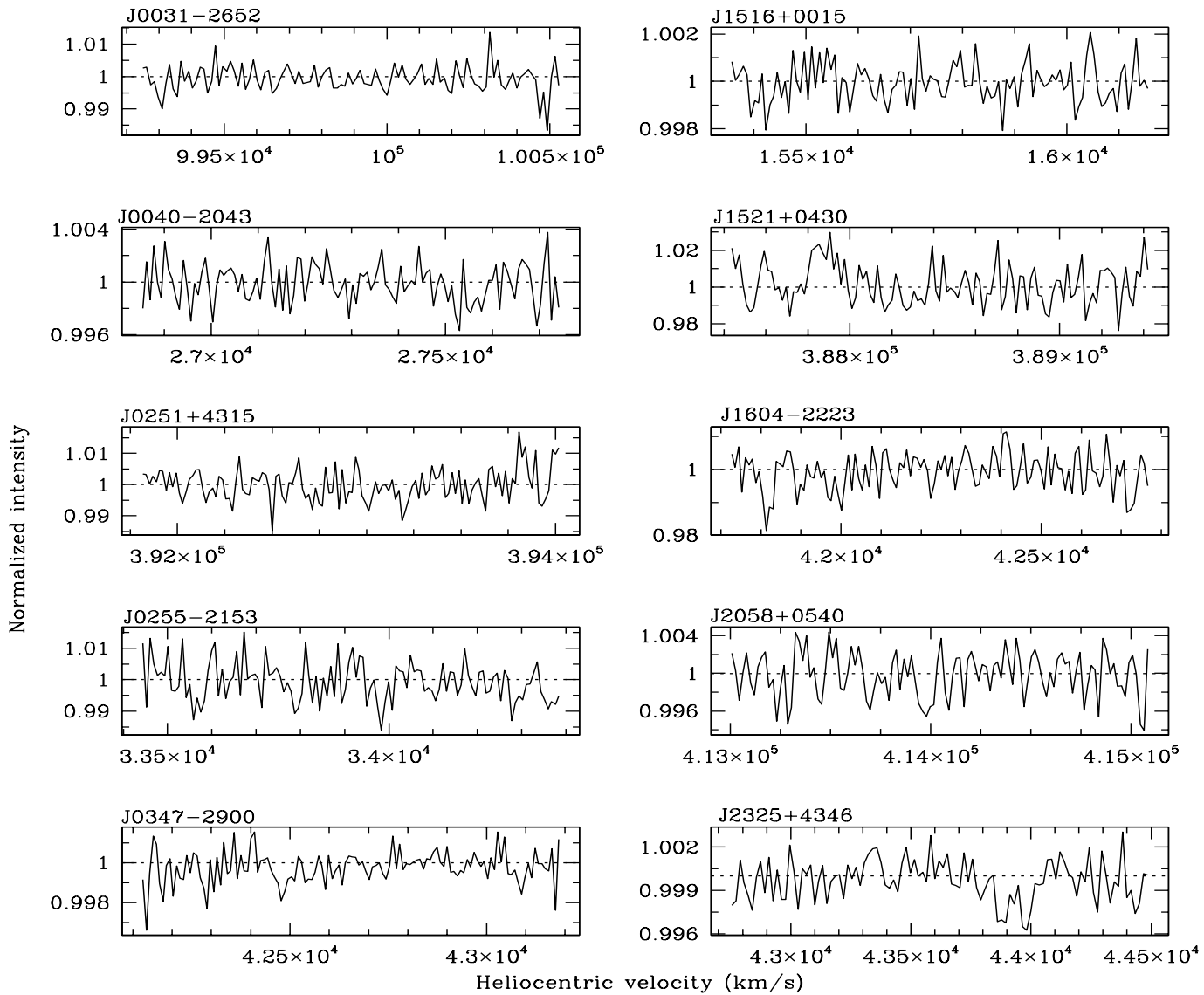,height=16.5cm,width=19.0cm,angle=0} }}
\caption[]{GMRT spectra for non-detections. The y-axis shows the normalized intensity while the x-axis shows the
heliocentric velocity in km s$^{-1}$.}
\label{fspec}
\end{figure*}
%
%%%%%%%%
The CSS quasar J0137+3309 (3C48) was observed on a number of occasions (see Table ~\ref{GMRTobs3C48}) 
to search for H{\sc i} absorption at $z_{abs}$=0.3654,
the redshift at which Gupta, Srianand \& Saikia (2005) detect outflowing material in the UV, and
$z$=0.3695 and 0.3700, the emission-line redshifts from CO and optical observations respectively (cf. Gupta et al. 2005).
The corresponding redshifted frequencies for 21-cm H{\sc i} absorption lie in the 1060-MHz sub-band of GMRT.
RFI at these frequencies not only makes the unambiguous detection of absorption and emission lines difficult,
but also affects bandpass stability, thereby reducing the spectral dynamic range.  Although a BB BW of 8
MHz was needed to cover the required redshifted frequencies,  we chose to use a BB BW of 4 MHz and move this in
frequency for the different observing sessions.  Since 3C48 is itself a standard calibrator, it was not
required to observe a separate flux density or phase calibrator.  However, the high flux density of 3C48
($\sim$ 20 Jy at 1060 MHz) demands observing an equally strong bandpass calibrator, and for similar durations.
We planned our observations such that 3C48 and 3C147 were observable at the same time.  In all observing runs on
3C48, we observed it and 3C147 with a duty cycle of about 30 min.  A log of the 3C48 observations is presented
in Table~\ref{GMRTobs3C48}.  The integration time in column 3 includes the observing time for 3C147.

%%%%%
The NRAO AIPS package was used for the GMRT data reduction.  In general, the data of the flux density calibrator,
phase calibrator and source were examined at various stages for bad antennas, baselines and time ranges.
In particular, the AIPS task SPFLG was invaluable for identifying data affected by RFI.  Such data were then flagged
out.  The complex gains determined from the final cleaned data on the flux density, bandpass, and phase calibrators
were used to calibrate the observed visibilties of each program source.  A continuum image of a source was then made
via the task IMAGR using data averaged over 10$-$15 line-free channels on both sides of the line.
This image was then self-calibrated until a satisfactory map was obtained.  This was not difficult as most fields
are dominated by a strong point source at the phase centre.  This continuum image was then Fourier transformed
and subtracted from the visibilty data cube using the task UVSUB.  The continuum-subtracted data cube was then imaged
(without cleaning) for all 128 channels to obtain a final image cube.  The spectrum was obtained by taking a cut
across the cube at the pixel correponding to the source location. If necessary, a smooth low-order
(usually order one) cubic spline was fitted to take out residual bandpass effects or continuum subtraction errors.
The spectra were then corrected to the heliocentric standard of rest.

%%%%%%%%%%%%%%%%%%%%%%%%%%%%%%%%%%%
\section{Observational results}
%%%%%%%%%%%%%%%%%%%%%%%%%%%%%%%%%%%
\subsection{H{\sc i} observations}
%%%%%%%%%%%%%%%%%%%%%%%%%%%%%%%%%%

From our search for H{\sc i} absorption towards 27 radio sources (Table~\ref{sample}), we made one new detection
(3C258) and confirmed 5 previously reported H{\sc i} absorption systems.  In the case of NGC\,1167, we have
confirmed the double-horn shaped H{\sc i} emission profile associated with the host E-S0 galaxy, which was
reported by Wegner, Haynes \& Giovanelli (1993).  Of the remaining 20 sources, there was no useful data for
J1643+1715 (3C346), we have not been able to verify the detection towards J0901+2901 (3C213.1) by Vermeulen et al.
(2003) while for J2250+1419 (4C+14.82) our Arecibo spectrum provides significant improvement over the limit
of Vermeulen et al. For J0137+3309 (3C48), we do not detect any absorption line in our deeper spectrum
although a preliminary detection was reported by Pihlstr\"om (2001).    The N(H{\sc i}) upper limits on
the remaining 16 sources are new.

The absorption lines detected were fitted with multiple Gaussians to determine the peak optical depth ($\tau_p$)
and FWHM ($\Delta v$; in \kms) of spectral components.  The parameters derived from the fits are considered in
detail in Section~\ref{source_notes}.  H{\sc i} column densities were determined using the relations
\begin{eqnarray}
{\rm N(HI)} & = & 1.835\times 10^{18}\frac{{\rm T_s}\int\!\tau (v)\,dv}{f_c} \, cm^{-2}  \nonumber \\
  & = &  1.93\times10^{18}\frac{{\rm T_s}\tau_p\Delta v}{f_c}  \, cm^{-2}
\label{eqcol}
\end{eqnarray}
where T${\rm _s}$ and $f_c$ are the spin temperature (in K) and the fraction of the background emission covered by
the absorber respectively. We have assumed T${\rm _s}$=100 K and $f_c$=1.0. For non-detections, the upper
limits on the H{\sc i} column densities were calculated by replacing the $\tau_p$ in the above equation
by 3$\sigma$ upper limits to the peak optical depths estimated from the rms in the spectra, and
assuming $\Delta v$=100 \kms.
Measured H{\sc i} column densities, and limits estimated as above, for the Arecibo sources are contained
in Table.~\ref{obsArres}.
The rms noises in the GMRT spectra, and constraints on the peak H{\sc i}
optical depths and column densities towards the GMRT sources, are presented in the last 3
columns of Table~\ref{GMRTobs+res}; the spectra for GMRT non-detections are shown in Fig.~\ref{fspec}.
One of the deepest Arecibo non-detections is shown in Fig.~\ref{J0901}.
The spectra for all detections are presented in Section~\ref{source_notes}.

%%%%%%%%%%%%%%%%%%%
\subsection {OH observations}
%%%%%%%%%%%%%%%%%%%%%%%%%
The 15 sources observed with the Arecibo telescope were also searched for OH main-line emission and/or absorption.
The observations were unaffected by RFI for 9 objects, but none revealed either emission or absorption.
Interpretation of these results is presented in Section~\ref{ohimp}.

%%%%%%%%%%%%%%%%%%%%%%%%%%%%%%%%%
\subsection{Notes on individual sources}
\label{source_notes}
%%
%
%%%%%%
{\bf J0034+3025:} This galaxy was not detected in either H{\sc i} or OH.  For OH, the segment of the
observed spectrum between 1420.14 and 1420.72 MHz was corrupted by the presence of Galactic H{\sc i}.

\begin{figure}
\centerline{\vbox{
\psfig{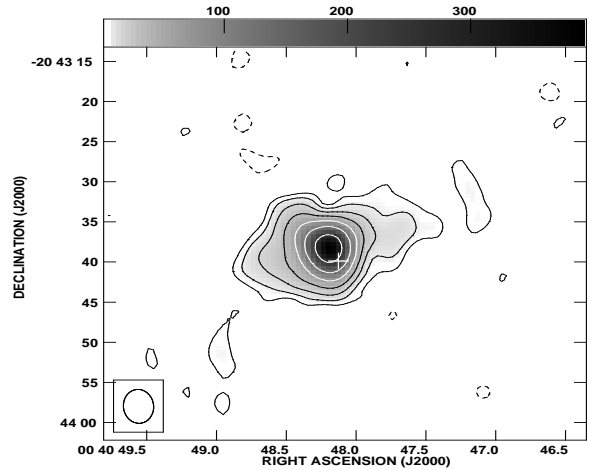}
}}
\caption[]{GMRT image of J0040$-$2043 at 1300 MHz with an rms noise of 1.4 mJy/beam.  The contour levels are 4$\times$(
$-$1, 1, 2, 4, 8, 16, 32 and 64) mJy/beam.  The restoring beam (see Table~\ref{GMRTobs+res}) is shown
as an ellipse and the position of the optical host galaxy (McCarthy et al. 1996) is marked with a cross.}
\label{0040sor}
\end{figure}

%%%%%%
\noindent
{\bf J0040$-$2043:}  This source was listed earlier as unresolved with an LAS$<$5$^{\prime\prime}$
(Kapahi et al. 1998a).  Our GMRT image (Fig.~\ref{0040sor}) at $\sim$1300 MHz shows an unresolved component
of LAS$\lapp$2.4$^{\prime\prime}$ and diffuse emission extending up to at least 15$^{\prime\prime}$.
From our map, the peak brightness and integrated flux density are 0.40 Jy/beam and 0.61 Jy respectively.

%%%%%
\noindent
{\bf J0119+3210 (4C+31.04):}
The mas-scale radio images of this source show two lobes separated by $\sim$80 mas along a  position
angle (PA) of $\sim$110$^\circ$ (Cotton et al. 1995).  A faint flat-spectrum component was identified by
Giovannini et al. (2001) as the core which, with the absence of any jet-like structures, led them to suggest
that the source is oriented close to the plane of the sky, with an angle of inclination to the
line of sight, $\theta\ge$ 75$^\circ$.

H{\sc i} absorption was first detected in this object by van Gorkom et al. (1989) and Mirabel (1990), who
discovered that the source also displayed a high velocity component relative to the systemic velocity
of the galaxy. The full H{\sc i} spectrum was subsequently studied in detail via VLBI observations by Conway (1999).
He interpreted the distribution of H{\sc i} opacity as due to an H{\sc i} disk whose axis is aligned with the
jet axis, i.e. close to the plane of the sky.  The disk covers the eastern lobe completely, but only the inner
part of the western lobe.  The high velocity component then arises from small clouds evaporating off the
inner edge of the disk.  The existence of this disk is further supported by an S-shaped dust lane with an
inner PA of $\sim$25$^\circ$ as seen in HST images (Capetti et al. 2000).

%%
%%%%%%%%%%%%%%%%%%%%%%%
%
\begin{figure}
\centerline{\vbox{
\vspace{-2.3cm}
\psfig{figure=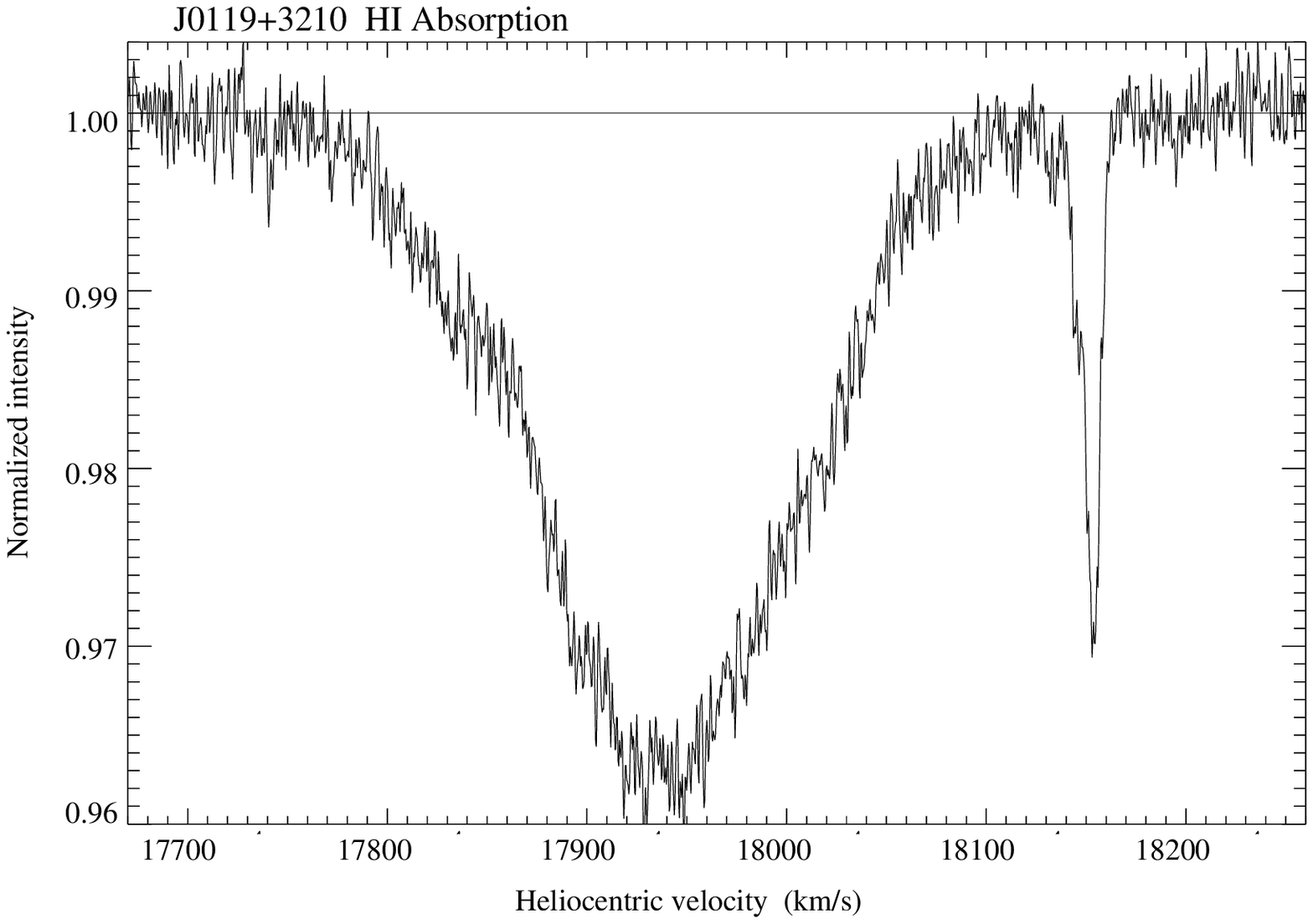,height=11cm,width=13.5cm,angle=0}
\vspace{-4cm}
\psfig{figure=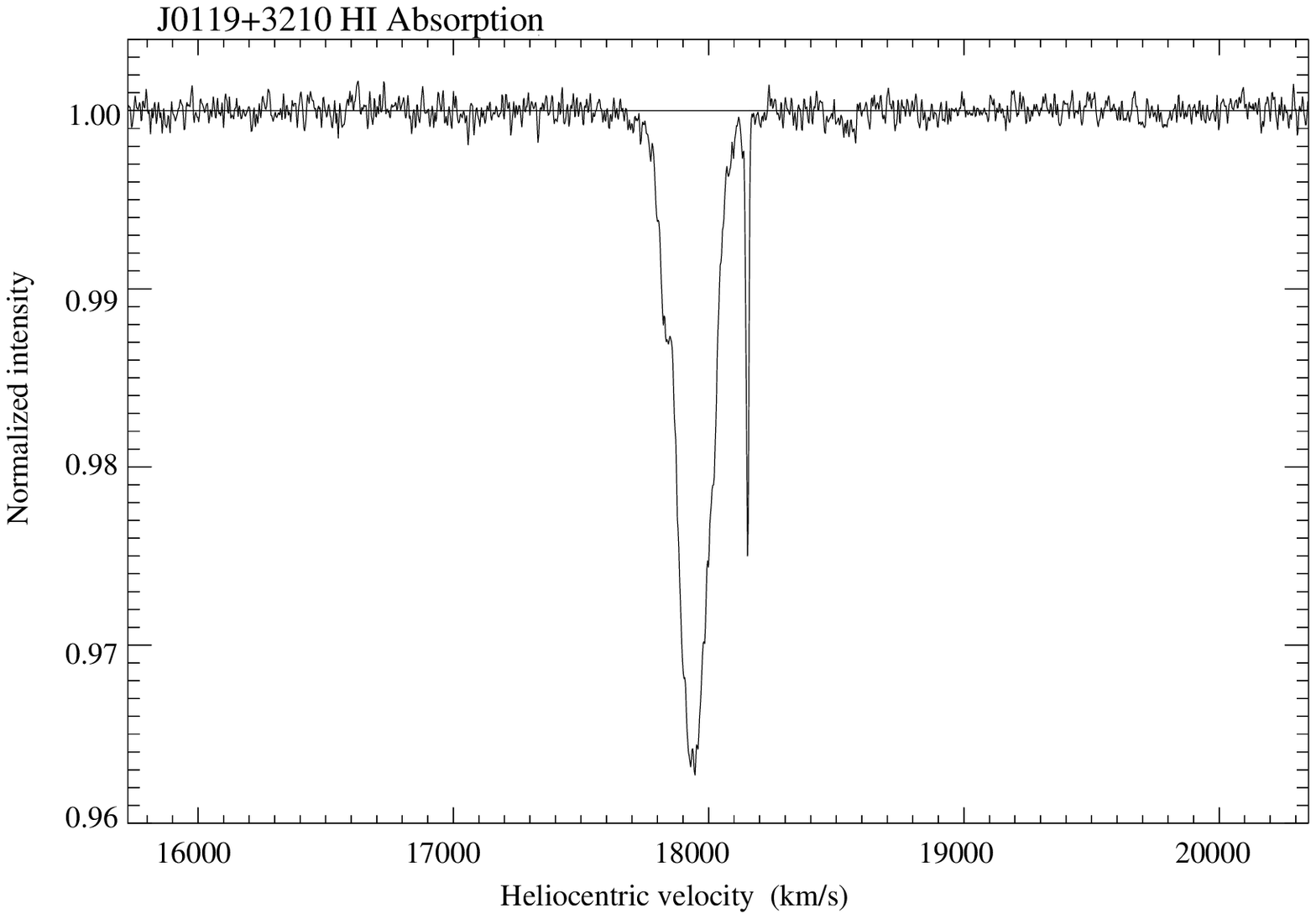,height=11cm,width=13.5cm,angle=0}
}}
\vspace{-2cm}
\caption[]{H{\sc i} absorption spectra of J0119+3210 (4C+31.04). Top: The high-resolution
spectrum.  Bottom: The low-resolution spectrum with the wider wavelength coverage.
}
\label{J0119+3210}
\end{figure}
\begin{figure}
\centerline{\vbox{
\psfig{figure=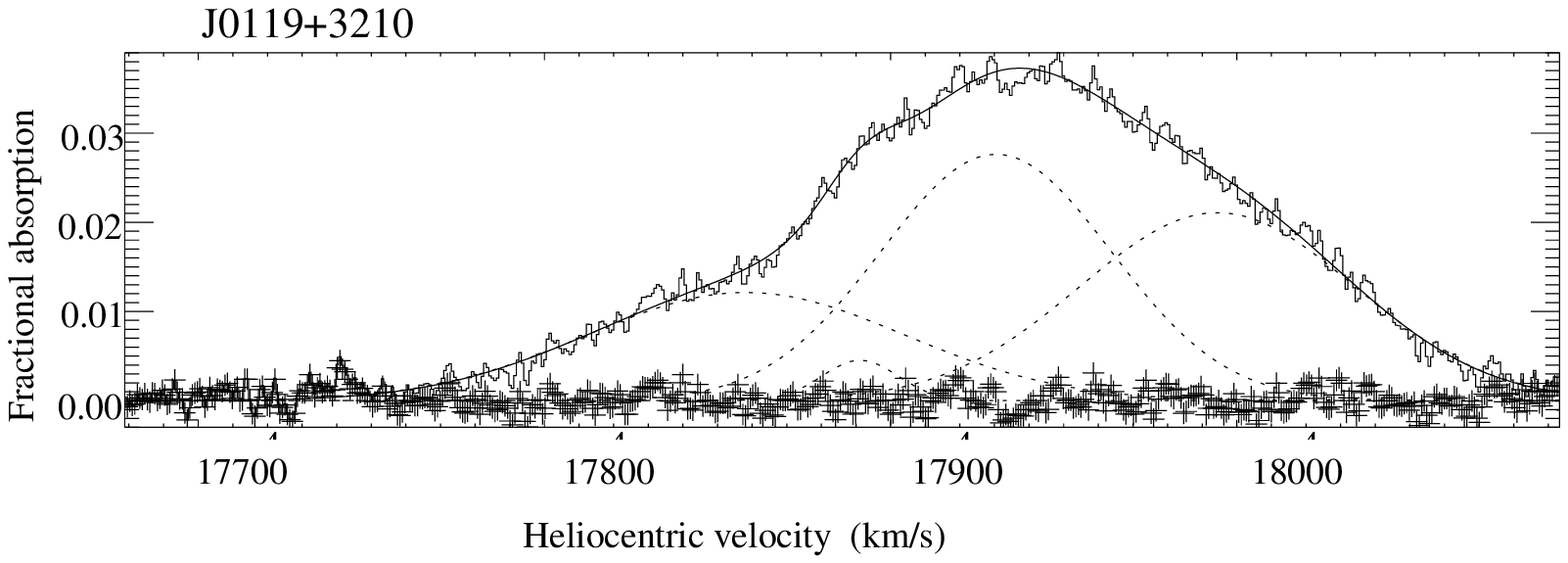,bbllx=18pt,bblly=7pt,bburx=488pt,bbury=169pt,width=8.45cm,height=3.5cm,angle=0}
\vspace{3.4cm}
\hspace{-2.8cm}
\psfig{figure=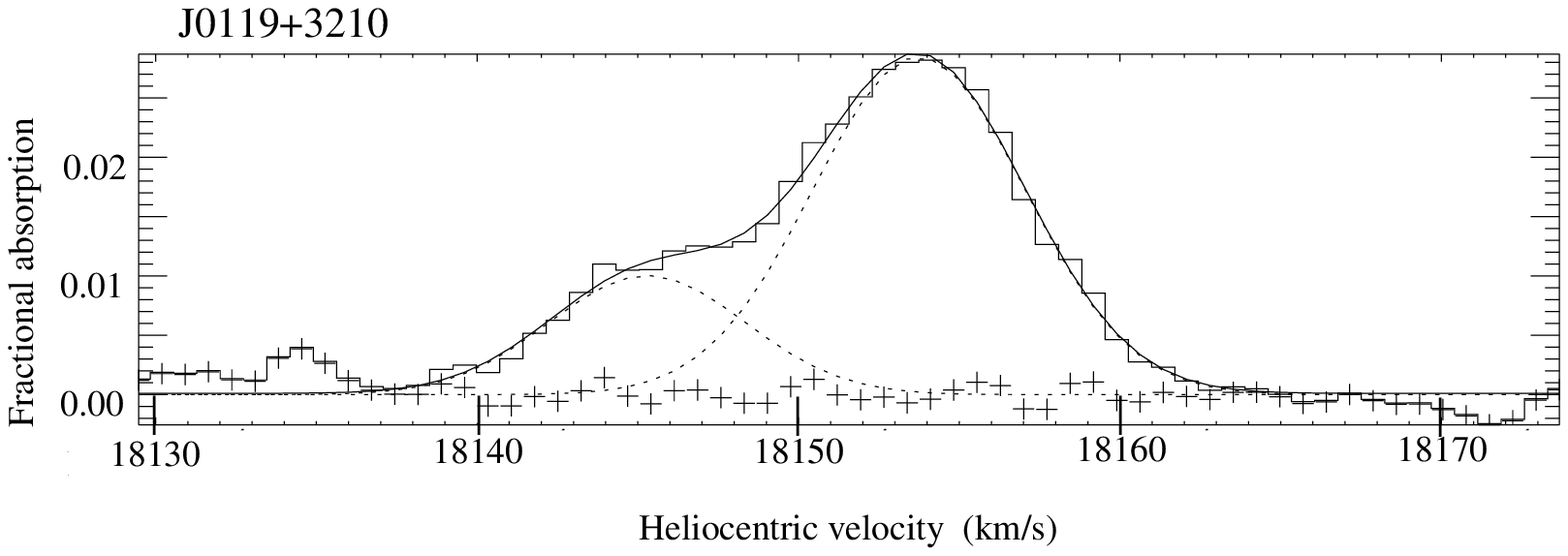,bbllx=13pt,bblly=180pt,bburx=488pt,bbury=340pt,width=8.45cm,height=3.5cm,angle=0}
}}
\vspace{-2.5cm}
\caption[]{The Gaussian fits to the H{\sc i} absorption spectrum of J0119+3210 (4C+31.04).  The high-resolution
spectrum (histogram) with a velocity resolution of 0.72 \kms ~is plotted along with the four-component Gaussian
fit to the broad component (top) and two-component Gaussian fit to the narrow high-velocity component (bottom).
The individual Gaussians are shown dotted, and their sum as a solid line. The residuals, following subtraction of
the Gaussians, are shown as crosses.
}
\label{J0119+3210gfits}
\end{figure}
%

%%%%%%%%%%%%%%%%%%%%
Sadly, the OH main lines are redshifted near to the frequency of GPS-L1 transmissions making them
unobservable.  Consequently, we decided to observe the H{\sc i} spectrum at Arecibo with both high and low
resolution, using total bandwidths of 3.125 and 25.0 MHz, each with 2048 spectral channels. The high resolution
spectrum (Fig.~\ref{J0119+3210}: top) was used to fit Gaussian components to the main and high-velocity
feature separately (Fig.~\ref{J0119+3210gfits}), while the low resolution spectrum (Fig.~\ref{J0119+3210}:
bottom) was used to search for any further weak absorption components separated in velocity from the others.
The number of Gaussians fitted here, (and for other sources), was the minimum needed to reduce the post-fit
residuals to noise.

For the high resolution spectrum, the main component is fitted well by four Gaussians, while the high-velocity
feature requires only two. These are summarized in Table~\ref{J0119fit}.
The low-resolution spectrum shows only a single possible absorption feature separated from the stronger features;
a small dip of optical depth $\sim$0.001 at heliocentric velocity $\sim$ 18530 km s$^{-1}$. This appears at the
3.5$\sigma$ level on a smoothed version of Fig.~\ref{J0119+3210}: bottom.  Confirmation would be required before
its reality could be claimed with confidence.

\begin{table}
\caption{Multiple-Gaussian fits to the broad and narrow absorption features
towards the radio source J0119+3210 (4C+31.04).}
\begin{center}
\begin{tabular}{l|l|l|c|c|}
\hline
Id. & v$_{\rm{hel}}$ & FWHM & Frac. abs. & N(H{\sc i})               \\
no. & \kms           &  \kms&            &  10$^{20}$cm$^{-2}$  \\
\hline
\multicolumn{5}{c}{Broad} \\
%Broad \\
1 & 17858.2(3.3) & 108.2(2.9)& 0.0121(0.0001) &  2.53(0.07)\\
2 & 17891.2(0.7) & 26.1(2.2) & 0.0045(0.0001) &  0.23(0.02) \\
3 & 17930.5(0.4) & 78.3(0.7) & 0.0276(0.0003) &  4.17(0.06) \\
4 & 17994.4(1.1) & 93.7(0.9) & 0.0211(0.0001) &  3.81(0.04) \\
\hline
\multicolumn{5}{c}{Narrow} \\
%Narrow \\
1 & 18145.3(0.06) & 7.19(0.09) & 0.0100(0.0002) &0.14(0.01) \\
2 & 18153.7(0.01) & 7.78(0.01) & 0.0285(0.0002) &0.43(0.01) \\
\hline
\end{tabular}
\end{center}
\label{J0119fit}
\end{table}
%
%

\begin{figure}
\centerline{\vbox{
\psfig{figure=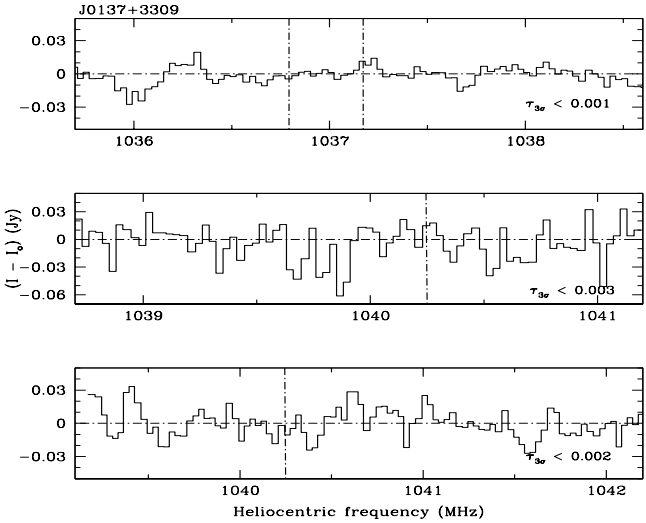,height=7.3cm,width=9.0cm,angle=0}
}}
\caption[]{GMRT spectra of J0137+3309 (3C48) covering the redshifted 21-cm frequency
(marked by vertical lines) corresponding to $z_{em}$=0.3700 from the optical emission lines, $z_{em}$=0.3695
from the CO emission spectrum (upper panel) and $z_{abs}$=0.3654 corresponding to the blue-shifted
absorbing material (middle and lower panels) detected in the UV by Gupta et al. (2005).
}
\label{spec3c48}
\end{figure}

%%%%%%
\noindent
{\bf J0137+3309 (3C48):}  This enigmatic CSS quasar has a complex radio morphology which has been
interpreted as being due to the interaction of the jet with the host galaxy's interstellar medium
(Wilkinson et al. 1991).  The jet-cloud interaction scenario is further supported by the
properties of blue-shifted gas detected in forbidden emission lines (Chatzicristou, Vanderriest \& Jaffe 1999).
Recently, Gupta et al. (2005) reported the detection of a $z_{abs}$=0.3654 associated absorption-line system in
an UV spectrum.  A tentative detection of H{\sc i} absorption towards 3C48 was reported by Pihlstr\"om
(2001) with a peak optical depth of 0.001 and FWHM of 100 km\,s$^{-1}$.  
Our deep GMRT observations (Fig.~\ref{spec3c48}) fail to 
reveal detectable H{\sc i} absorption at either
the systemic redshift of 3C48 or the redshift corresponding to the gas responsible for the UV absorption lines.
For the flux density of 20.3 Jy determined from our observations and assuming $f_c$=1.0, we obtain a
3$\sigma$ upper limit on optical depth in the range of 0.001 to 0.003 from the unsmoothed spectra shown in 
Fig.~\ref{spec3c48}.

\begin{figure}
\centerline{\vbox{
\psfig{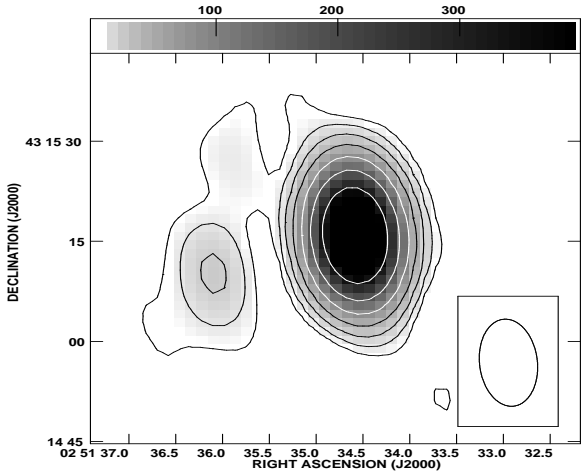}
}}
\caption[]{GMRT image of J0251+4315  at 616 MHz with an rms noise of 1.6 mJy/beam.  The contour levels are
5$\times$($-$2, $-$1, 1, 2, 4, 8, 16, 32 and 64) mJy/beam.  The restoring beam (see Table~\ref{GMRTobs+res}) is
shown as an ellipse. The eastern feature is the foreground galaxy G0248+430.
}
\label{0251sor}
\end{figure}

%%%%%%%%%%%%%%%%%
\noindent
{\bf J0251+4315:}
The source structure in the VLBA images at 2.3 and 8 GHz show two prominent components, but appear to be
better described by a multi-component structure (Fey \& Charlot 2000).  We detect no significant H{\sc i}
absorption feature associated with this GPS quasar in our GMRT spectrum.  The quasar is part of a quasar-galaxy
pair with the foreground galaxy G0248+430 ($z$=0.052) located at a distance of $\sim$15$^{\prime\prime}$ along
a PA of 105$^\circ$ (Sargent \& Steidel 1990).  The GMRT 616-MHz image (Fig.~\ref{0251sor}) also shows continuum
emission from the galaxy which has a flux density of $\sim$25 mJy at this frequency.

\begin{figure}
\centerline{\vbox{
\vspace{-2.3cm}
\psfig{figure=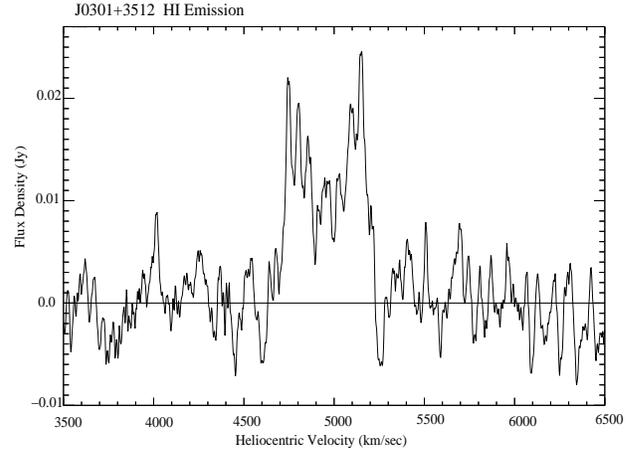,height=11cm,width=13.5cm,angle=-90}
}}
\vspace{-2.3cm}
\caption[]{The double-horn shaped H{\sc i} emission profile of the E-SO galaxy NGC 1167 (J0301+3512).}
\label{J0301+3512}
\end{figure}

%%%%%%
\noindent
{\bf J0301+3512 (NGC1167):}
The Arecibo H{\sc i} spectrum of this low redshift, E-S0, galaxy, which has a somewhat complex radio structure
(Sanghera et al. 1995), shows no recognizable absorption. However, an emission spectrum typical of a spiral
galaxy is seen (Fig.~\ref{J0301+3512}), where the spectrum has been normalized into Jy\, beam$^{-1}$,
as is appropriate for line emission. The parameters derived from the spectrum are a H{\sc i}
line flux density of 6.22$\pm$0.4(1$\sigma$) Jy\,km s$^{-1}$, a heliocentric radial velocity of
4965 km s$^{-1}$, and line widths of 504 and
511 km s$^{-1}$ at 50\% of the mean and 20\% of the peak signal intensity respectively.  These values agree
reasonably with those given by Wegner et al. (1993). As implied by the entry for J0301+3512 in Table~2
of Wegner et al., the H{\sc i} emission from this galaxy is expected to be significantly resolved by
the Arecibo beam at 21-cm.  Earlier H{\sc i} observations have been reported by Heckman et al. (1983) and
Chamaraux, Balkowski \& Fontanelli (1987) who suggested that the H{\sc i} gas distribution
may be influenced by tidal interactions with two disk galaxy companions.

%%%%%%
\noindent
{\bf J0513+0157 (4C+01.13):}
Perlman et al. (1998) classified the parent optical object as a BL Lac on the basis of Ca~{\sc ii} break
strength and lack of emission lines.  However, due to the low signal-to-noise ratio of the spectrum blueward
of the Ca~{\sc ii} break, its classification as a radio galaxy cannot be ruled out.  The OH observations for
this source were corrupted by extensive RFI across the band. At the time of observations, the best available
redshift for this source was $z=0.084$. However, Rines et al. (2003) have since measured an improved
value of $z=0.08808\pm0.00014$, which would place the H{\sc i} line towards the edge of the present spectrum.
Nevertheless, there is still sufficient baseline to say with confidence that no H{\sc i}
absorption is seen for the source at the 3$\sigma$ level for the parameters given in Table~\ref{obsArres}.

%%%%%%%
\noindent
{\bf J0725$-$0054:}
This narrow-line radio galaxy, which has an asymmetric radio structure with a knotty jet extending for
$\gapp$30 mas along PA$\sim-24^\circ$ (Fey \& Charlot 1997; Bondi et al. 1996), was observed with the
then best available redshift of $z = 0.128$. An improved value of $z=0.1273\pm0.0001$ has since been measured
by Eracleous \& Halpern (2004). RFI is present over part of the Arecibo H{\sc i} spectrum, with no H{\sc i}
absorption being detected in the uninterferred spectral region corresponding to $0.1247 < z < 0.1291$.
The OH spectrum is RFI-free, and again neither emission nor absorption is seen.

%%%%%%
\noindent
{\bf J0805+2409 (3C192):}
This FRII source has an X-shaped structure (Baum et al. 1988). It is marginally resolved by the Arecibo beam at
1340 MHz, its measured flux density representing $\sim$84\% of the total.  No H{\sc i} absorption was detected,
while the OH spectrum was destroyed by RFI from GPS-L1 transmissions.

%%%%%%
\noindent
{\bf J0822+0557 (3C198):}
The OH spectrum for this classical double-lobed FRII radio galaxy (Fomalont 1971), was destroyed by RFI.
With its large angular diameter, it is heavily resolved by the Arecibo beam at 1315 MHz, the measured flux
density representing only $\sim$63\% of the total. No H{\sc i} absorption was detected.

\begin{figure}
\centerline{\vbox{
%\vspace{-2.3cm}
\psfig{figure=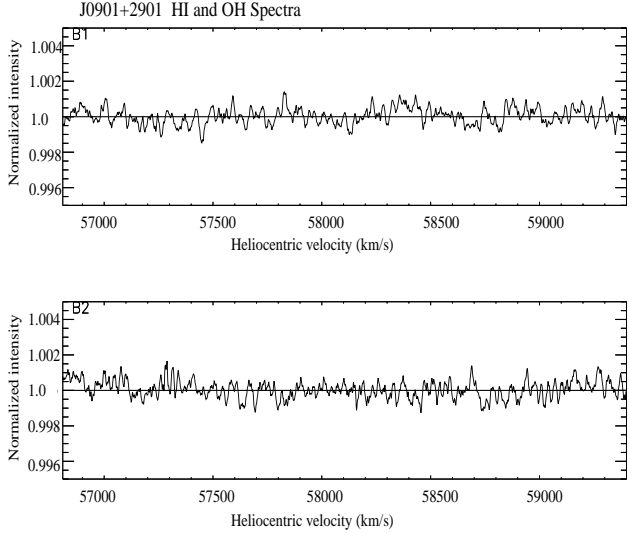,height=7.3cm,width=9cm,angle=0}}}
%\vspace{-2.3cm}
\caption[]{H{\sc i} and OH spectra for J0901+2901 (3C213.1); the y-axis is the spectral intensity normalized
by the source flux density, while the x-axis is the heliocentric velocity in km s$^{-1}$. Top: the H{\sc i} spectrum
with a velocity resolution of 16.5 km s$^{-1}$; bottom; the OH spectrum with a velocity resolution of 14.1 km s$^{-1}$.}
\label{J0901}
\end{figure}

%%%%%%
\noindent
{\bf J0901+2901 (3C213.1):}
This galaxy has emission that extends up to $\sim$40$^{\prime\prime}$ in addition to a more compact
double-lobed structure (Akujor et al. 1991).  A detection of H{\sc i} absorption in it was recently reported
by Vermeulen et al. (2003). The optical depth that they measured was $\tau = 0.0005$, the lowest value detected
in their survey by a factor of $\sim$3. Although this source provided one of the lowest rms optical depth
limits among our Arecibo observations (Fig.~\ref{J0901}:\,top), this is still not deep enough to verify the
detection of Vermeulen et al. The Arecibo OH spectrum (Fig.~\ref{J0901}:\,bottom) shows neither emission nor
absorption to a similar optical depth limit as for H{\sc i}. The spectra shown illustrate one of the deeper
Arecibo non-detections.

%%%%%
%%%%

%%%%
\begin{figure*}
\vspace{-2.3cm}
\centerline{\hbox{
\hspace{-1.0cm}
\psfig{figure=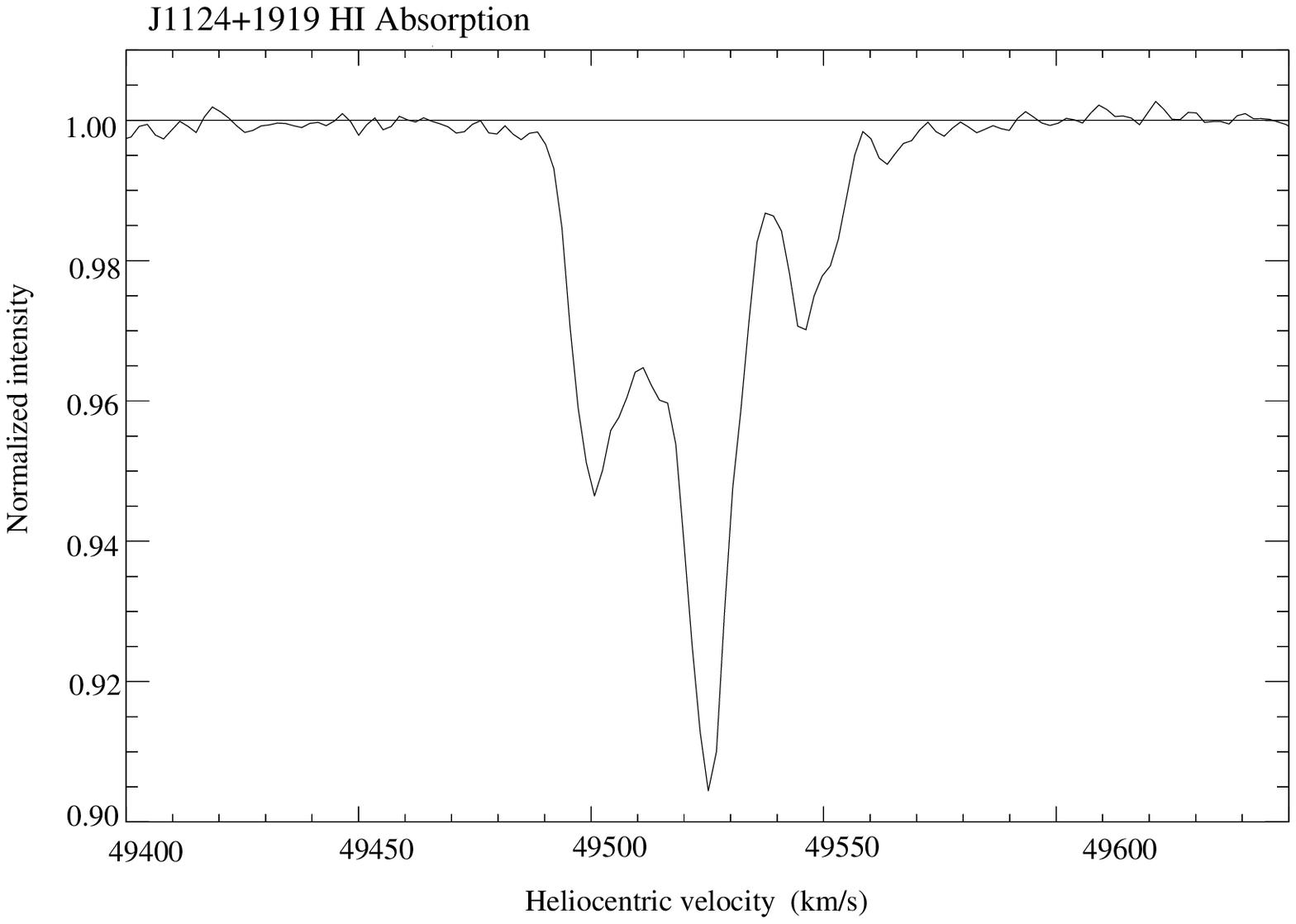,height=11cm,width=13.5cm,angle=-90}
\hspace{-4.5cm}
\psfig{figure=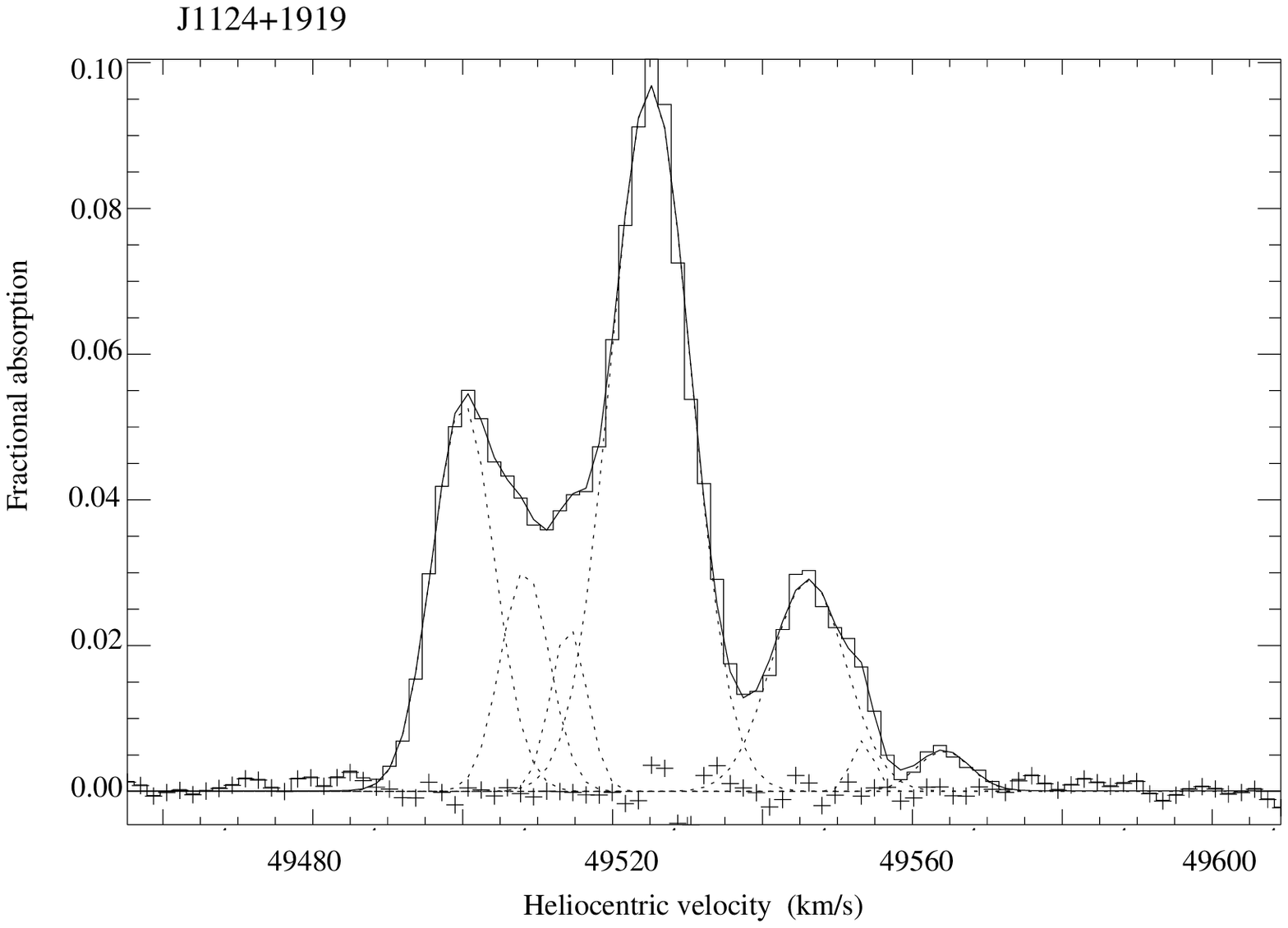,height=11cm,width=13.5cm,angle=-90}
}}
\vspace{-2.3cm}
\caption[]{Left: The H{\sc i} absorption spectrum for J1124+1919 (3C258) with a velocity resolution of 3.5 km s$^{-1}$.
Right: The same spectrum  plotted as a histogram, overlaid by the seven Gaussians fitted to the spectrum.
The meanings of the different line types are as in Fig.~\ref{J0119+3210gfits}.
}
\label{c258a}
\end{figure*}

%%%%%%%
\noindent
{\bf J1124+1919 (3C258):}
Deep H{\sc i} absorption (Fig.~\ref{c258a}:\,left) was detected for the first time against the continuum
emission from this radio galaxy. This emission is dominated by the CSS source but it is known to also have more
extended emission (Strom et al. 1990; Sanghera et al. 1995). The FWHM of the line is at the narrow end when
compared to values listed by Pihlstr\"{o}m et al. (2003). Despite this, it shows considerable structure, and
seven Gaussian components are needed to fit the spectrum satisfactorily (Fig.~\ref{c258a}:\,right). These
components are listed in Table~\ref{c258gau}.

%%%%%%
\begin{table}
\caption{Multiple-Gaussian fit to the H{\sc i} absorption
spectrum of J1124+1919 (3C258).}
\begin{center}
%\vspace{5mm}
\begin{tabular}{|c|l|l|c|c|}
\hline
Id. & ~~~~~v$_{\rm{hel}}$ & ~FWHM & Frac. abs. &  N(H{\sc i}) \\
no.  & ~~~~km s$^{-1}$ & ~~km s$^{-1}$ & & 10$^{20}$\,cm$^{-2}$ \\
\hline
1 & 49500.1(0.05) & 9.69(0.04) & 0.0532(0.0013) & 1.00(0.03) \\
2 & 49508.3(0.09) & 7.64(0.28) & 0.0304(0.0011) & 0.45(0.02) \\
3 & 49514.2(0.07) & 5.71(0.14) & 0.0226(0.0026) & 0.25(0.03) \\
4 & 49525.1(0.03) & 12.50(0.01)& 0.0968(0.0002) & 2.34(0.01) \\
5 & 49546.1(0.06) & 11.73(0.11)& 0.0290(0.0002) & 0.66(0.01) \\
6 & 49553.4(0.20) & 3.60(0.49) & 0.0070(0.0008) & 0.05(0.01) \\
7 & 49564.2(0.38) & 8.23(0.78) & 0.0056(0.0002) & 0.09(0.01) \\
\hline
\end{tabular}
\end{center}
\label{c258gau}
\end{table}
%%%

%%%
The individual Gaussian components are narrow, with half-height widths lying between $\sim$4 and 12 km s$^{-1}$.
These narrow widths for the fitted components imply rather low upper limits on $\rm T_s$ values for the H{\sc i},
the derived values lying in the range T${\rm _s} \la 400-2000\,$K.  Compared to other CSS H{\sc i}-absorption
systems, 3C258 exhibits an unusually complex spectrum, with the components having about an order of magnitude
narrower velocity widths than for most, though not all, features in other CSS sources
(Vermeulen et al. 2003; Pihlstr\"{o}m et al. 2003). The possibility of the existence of weak, extended line
wings, as suggested by Fig.~\ref{c258a}:\,left, will have to await deeper integrations for confirmation.

%%%%
Although detailed modelling of the H{\sc i} gas seen in absorption in this galaxy will require VLBI imaging of
the line and continuum emission, it is difficult to imagine that a single rotating disk/torus would produce such
a complex absorption system.  This may be a miniature version of 3C236, where jet-ISM interactions play the
dominant role in shaping the structure/properties of the object.  In fact, the signature of disturbance is
perhaps already evident in its optical image.  The HST image of 3C258 (de Vries et al. 1997) displays an arc-like
structure  of high surface brightness, with a larger, fainter tail extending to the northeast, roughly perpendicular
to the bright central arc. The radio continuum image (Akujor et al. 1991) is at right angles
to the central region and contains the entire VLBI-scale double-lobed structure (Sanghera et al. 1995).
Neither OH absorption nor emission were detected against 3C258.

\begin{figure}
\centerline{\vbox{
\psfig{figure=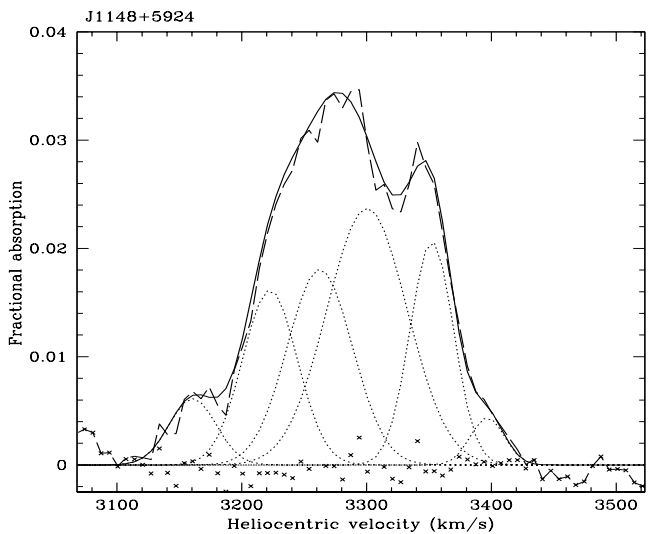,height=7.3cm,width=9.0cm,angle=0}
}}
\caption[]{
The H{\sc i} absorption spectra (dashed line) for J1148+5924 (NGC3894) obtained with the GMRT.
The spectrum has been smoothed using a 5 pixel wide boxcar filter.
The six Gaussian components fitted to the absorption profile, the sum of these component fits and
residuals are plotted as dotted and solid lines, and crosses respectively.
}
\label{specngc3894}
\end{figure}

%%%%%%
\noindent
{\bf J1148+5924 (NGC3894):}
The galaxy NGC3894, classified as an E/S0, lies at a redshift of z=0.01075 (de Vaucouleurs et al. 1991).
This is in agreement within the error of $\pm$30\,\kms, with the redshift of z=0.01068 determined by
Karachentsev (1980).  Using multi-epoch VLBI observations, Taylor, Wrobel \& Vermeulen  (1998) suggest that
the radio source is oriented well away from the line of sight ($\theta\sim$50$^\circ$).
Redshifted H{\sc i} absorption has been detected towards both the approaching and receding jets by
Peck \& Taylor (1998).  Our GMRT spectrum (Fig.~\ref{specngc3894}) shows the H{\sc i} absorption
towards this radio source.  A good fit to the absorption profile was obtained using six Gaussian components.
The best-fit parameters are summarised in Table~\ref{fitngc3894}.  Most of the absorption features are
redshifted with respect to the systemic velocity ($\sim$3220\,\kms) and could represent the large-scale
circumnuclear disk, as well as clouds infalling towards the central engine (Peck \& Taylor 1998).

%%%%%%%
%%%
%
\begin{table}
\caption[]{Multiple-Gaussian fit to the H{\sc i} absorption spectrum of J1148+5924 (NGC3894).}
\begin{center}
%\vspace{5mm}
\begin{tabular}{|c|l|l|c|c|}
\hline
Id. & v$_{\rm{hel}}$ & FWHM & Frac. abs. & N(H{\sc i}) \\
no. & km s$^{-1}$ & km s$^{-1}$ &         & 10$^{20}$cm$^{-2}$\\
\hline
1 &3161(4)   &   44.53(6.50)     & 0.0061(0.0006)      &  0.52(0.09)         \\
2 &3223(2)   &   51.74(3.58)     & 0.0161(0.0046)      &  1.61(0.47)         \\
3 &3262(2)   &   61.44(17.44)    & 0.0181(0.0024)      &  2.15(0.67)         \\
4 &3300(2)   &   75.77(12.26)    & 0.0237(0.0039)      &  3.47(0.80)         \\
5 &3352(2)   &   41.04(2.88)     & 0.0207(0.0024)      &  1.64(0.22)         \\
6 &3396(5)   &   32.26(7.34)     & 0.0044(0.0008)      &  0.27(0.08)         \\
\hline
\end{tabular}
\end{center}
\label{fitngc3894}
\end{table}
%%%
%%%

%%%%%%%%%%%%%%%%%%%%%
%%
\begin{figure}
\centerline{\vbox{
%\vspace{-2.3cm}
\psfig{figure=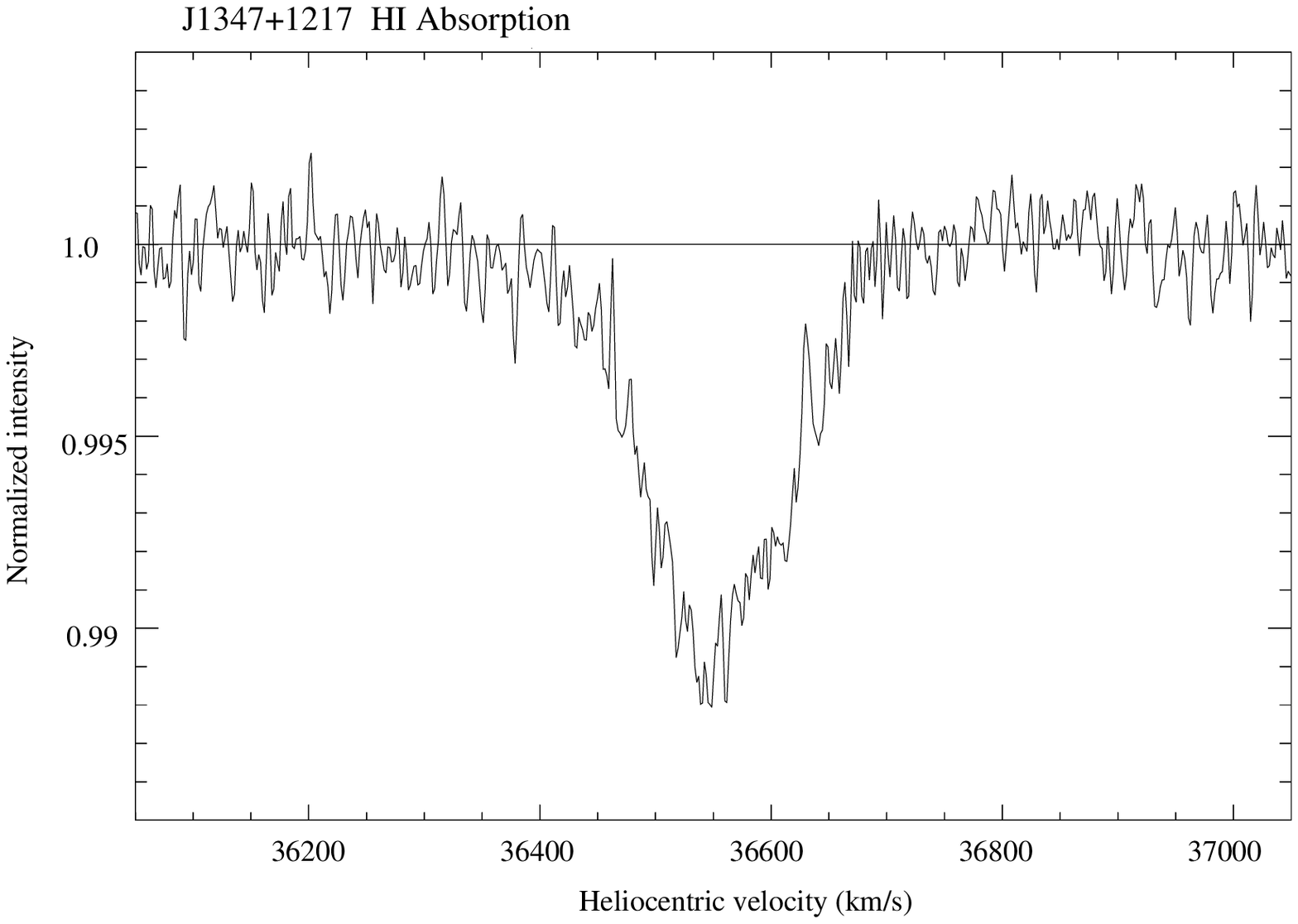,height=7.3cm,width=9cm,angle=0}
%\vspace{-4.5cm}
\psfig{figure=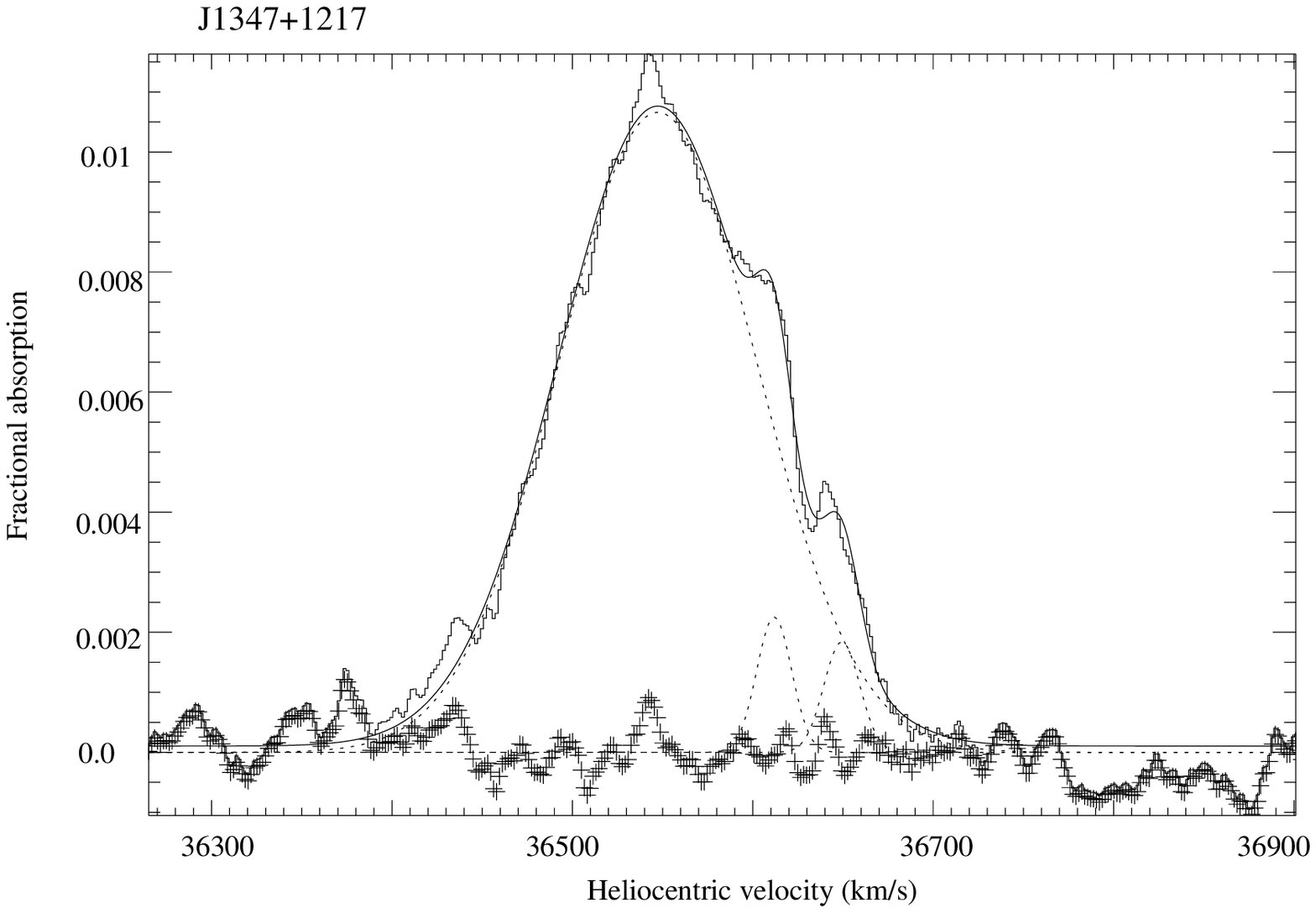,height=7.3cm,width=9cm,angle=0}
}}
%\vspace{-2cm}
\caption[]{Top: The H{\sc i} absorption spectrum towards the radio source J1347+1217 (4C+12.50)
at a spectral resolution of 3.2 km\,s$^{-1}$.
Bottom: The Gaussian fits to the components at a velocity resolution of 28.8 km\,s$^{-1}$.
The line types are as in Fig.~\ref{J0119+3210gfits}. }
\label{J1347+1217}
\end{figure}

%%%%%
\noindent
{\bf J1347+1217 (4C+12.50):}
This CSS source is associated with a Seyfert-2 galaxy containing two nuclei separated by 2$^{\prime\prime}$,
which may be in the process of merging (Gilmore \& Shaw 1986).  Broad Pa$\alpha$ emission
($\Delta v$$_{{\rm FWHM}}$$\approx$2600 \kms) suggests that it may contain a buried quasar (Veilleux, Sanders
\& Kim 1997). The radio structure is highly asymmetric and has a jet on VLBI scales
(Xiang et al. 2002; Lister et al. 2003).

Mirabel (1989) detected broad H{\sc i} absorption against this radio galaxy, followed shortly afterwards
by the detection (Mirabel, Sanders \& Kaz\`{e}s 1989) of very broad CO(1$\rightarrow$0) emission
(see also Scoville et al. 2000).  The H{\sc i} spectrum of Mirabel (1989) showed a deep `core'
near the redshift of the galaxy, and apparent very broad wings covering a total velocity range of 950 km s$^{-1}$.
Morganti et al. (2004) recently demonstrated that the line wings have an optical depth of $\la
0.002$, and cover some 2000 km s$^{-1}$, extending especially to the blue. We are unable to add anything on
these line wings, both because of sensitivity and the presence of RFI at 1370 MHz (v$ \sim 35500$
km s$^{-1}$). However, our spectrum has higher velocity resolution on the core than previously published examples,
and is shown in Fig.~\ref{J1347+1217}:top.  There is an indication of broadening of the redshifted side of the
line core, and a smoothed version of the spectrum can be fitted well by three Gaussian components
(see Fig~\ref{J1347+1217}:bottom), where the best-fit parameters are listed in Table~\ref{J1347fit}.

%%%%%
\begin{table}
\caption{Multiple-Gaussian fit to the H{\sc i} absorption spectrum of J1347+1217 (4C+12.50).}
\begin{center}
%\vspace{5mm}
\begin{tabular}{l|l|l|c|c|}
\hline
Id. & v$_{\rm{hel}}$ & FWHM   & Frac. abs. & N(H{\sc i}) \\
no. &  \kms          &  \kms  &            & 10$^{20}$cm$^{-2}$\\
\hline
1 & 36547.4(0.4) & 129.9(0.7) & 0.0107(0.0001) &  2.68(0.03) \\
2 & 36611.9(0.8) & 22.8(1.8)  & 0.0023(0.0001) &  0.10(0.01) \\
3 & 36649(1.7)   & 23.5(3.9)  & 0.0018(0.0001) &  0.08(0.01) \\
\hline
\end{tabular}
\end{center}
\label{J1347fit}
\end{table}
%%%%

%%%%%
A recently published redshift for this object by Holt, Tadhunter \& Morganti (2003) gives
$z$ = 0.12174 $\pm$ 0.00002 (v $= 36497 \pm 6$ km s$^{-1}$).  This is consistent with our broad core
component \#1 being associated with the nucleus of the galaxy.

From the OH spectrum for J1347+1217 neither emission nor absorption is detected. The source was included
in the sample of Darling \& Giovanelli (2002), for which they searched for OH megamaser emission at Arecibo.
However, they were unable to provide a meaningful limit to such emission for this object due to the effects
of standing waves.  Dickey et al. (1990) reported a tentative detection of OH emission in this source
of peak flux density $1.7\pm0.2$~mJy, and a line width of $305\pm44$~km s$^{-1}$.  Smoothing our spectrum to
a velocity resolution of 35~km s$^{-1}$, we can set an upper limit on the presence of 1667-MHz
OH emission of 4.8 mJy (3$\sigma$). A deeper integration would be valuable.

%%
%%%%%%
\begin{figure} \centerline{\vbox{
%\vspace{-2.3cm}
\psfig{figure=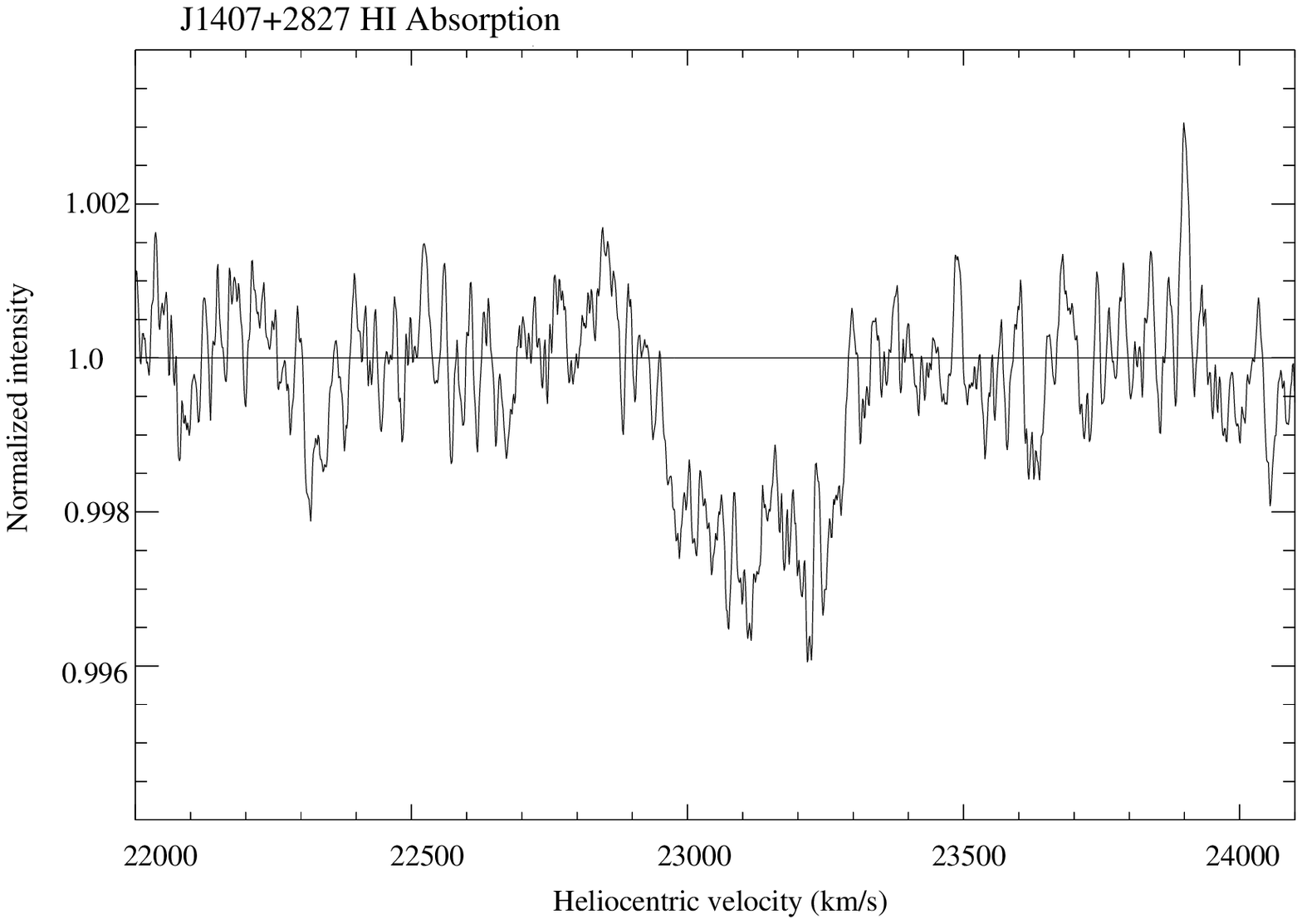,height=7.3cm,width=9.0cm,angle=0}
%\vspace{-4.5cm}
\psfig{figure=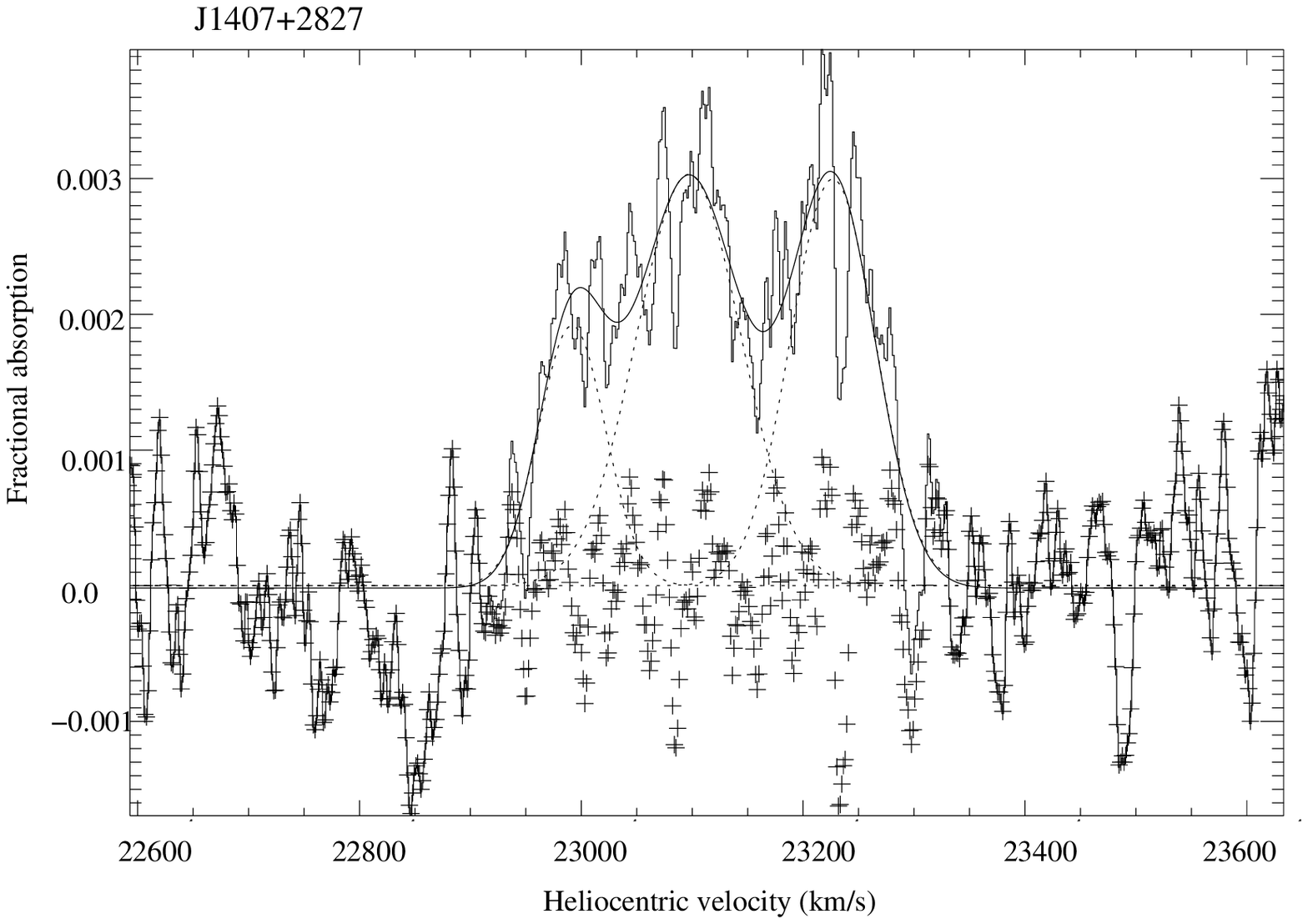,height=7.3cm,width=9.0cm,angle=0}
}}
%\vspace{-2.0cm}
\caption[]{Top: The H{\sc i} absorption spectrum towards the radio source J1407+2827 (Mrk668).
Bottom: The three Gaussian components are overlaid on the spectrum. The line types are as
in Fig.~\ref{J0119+3210gfits}.}
\label{J1407+284}
\end{figure}
\begin{table}
\caption[]{Multiple-Gaussian fit to the H{\sc i} absorption spectrum of J1407+2827 (Mrk668).}
\begin{center}
%\vspace{5mm}
\begin{tabular}{l|l|l|c|c|}
\hline
Id. & v$_{\rm{hel}}$ & FWHM & Frac. abs. & N(H{\sc i}) \\
no. &  \kms        &  \kms &           & 10$^{20}$cm$^{-2}$\\
\hline
1 & 22991(16) & 71(23) & 0.0019(0.0001) & 0.26(0.08)  \\
2 & 23097(7)  & 111(21)& 0.0030(0.0001) & 0.64(0.12)  \\
3 & 23227(6)  & 90(11) & 0.0030(0.0001) & 0.52(0.07)  \\
\hline
\end{tabular}
\end{center}
\label{J1407fit}
\end{table}
%

%%%%%%
\noindent
{\bf J1407+2827 (Mrk668):}
This GPS object has an asymmetric double-lobed structure and is embedded in a galaxy with highly irregular
isophotes, suggesting dynamical disturbance (Stanghellini et al. 1993, 2001).  Broad, but weak, H{\sc i}
absorption was detected against this source by Vermeulen et al. (2003). We confirm this (Fig.~\ref{J1407+284}),
and find evidence of structure within the feature. The line is well fitted by three Gaussian components,
the best-fit parameters being listed in Table~\ref{J1407fit}.

Recently published redshifts for this object are $z$=0.07658$\pm$0.00013 (Huchra et al. 1990) and
$z$=0.07681$\pm$0.00007 (Eracleous \& Halpern 2004). Our measured redshift for the fitted H{\sc i} components
of $z$ = 0.07669(5), 0.07704(2) and 0.07748(2) suggest that the two stronger components are consistent with
infall on to the galactic nucleus. Unfortunately, the OH spectrum for this source was corrupted by
extensive RFI.

%%%%%%
\begin{figure}
\centerline{\hbox{
  \psfig{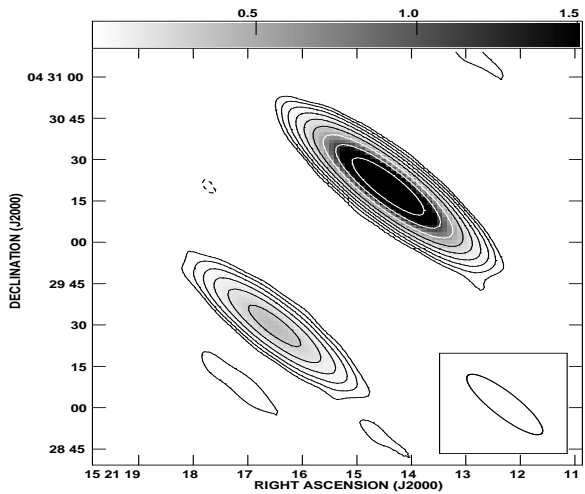}
}}
\caption[]{GMRT image of J1521+0430 (4C+04.51) at 619 MHz with an rms of 3 mJy/beam.  The contour levels are
9$\times$($-$2, $-$1, 1, 2, 4, 8, 16, 32, 64, 128 and 256) mJy/beam.  The restoring beam (see Table~\ref{GMRTobs+res}) is
shown as an ellipse.}
\label{1521sor}
\end{figure}
%

%%%%%%%
\noindent
{\bf J1521+0430 (4C+04.51):}
In addition to the compact VLBI-scale double of this GPS source, which is unresolved in our 619-MHz observations,
there is a separate component towards the south-east at an angular distance of $\sim$1$^{\prime}$ (Fig.~\ref{1521sor}).
Comparing this with the NVSS image yields a spectral index of $\sim$0.7 between $\sim$620 and 1400 MHz.
It is unclear whether this component is related to our source, although $\sim$10\% of GPS sources are known
to have radio emission beyond kpc scales (Stanghellini et al. 1990).  A more sensitive low-frequency image is
required to establish whether this feature is related to the GPS source.

%%%%%%
\noindent
{\bf J1643+1715 (3C346):}
The parent galaxy of 3C346 has a double nucleus, the compact, north-western component being coincident
with the radio core (Dey \& van Breugel 1994).  The source has a one-sided radio jet extending for
$\sim$2$^{\prime\prime}$ (Cotton et al. 1995), and diffuse extended emission on a scale of
$\sim$12$^{\prime\prime}$ (Akujor et al. 1995).  The one-sided jet suggests a small angle of
inclination to the line of sight.  The H{\sc i} spectrum for this source were corrupted by the effects
of GPS-L2 transmissions.  However, the OH spectrum is good, but neither emission nor absorption is detected.

%%%%%%
\begin{figure}
%\vspace{-2cm}
\centerline{\vbox{
\psfig{figure=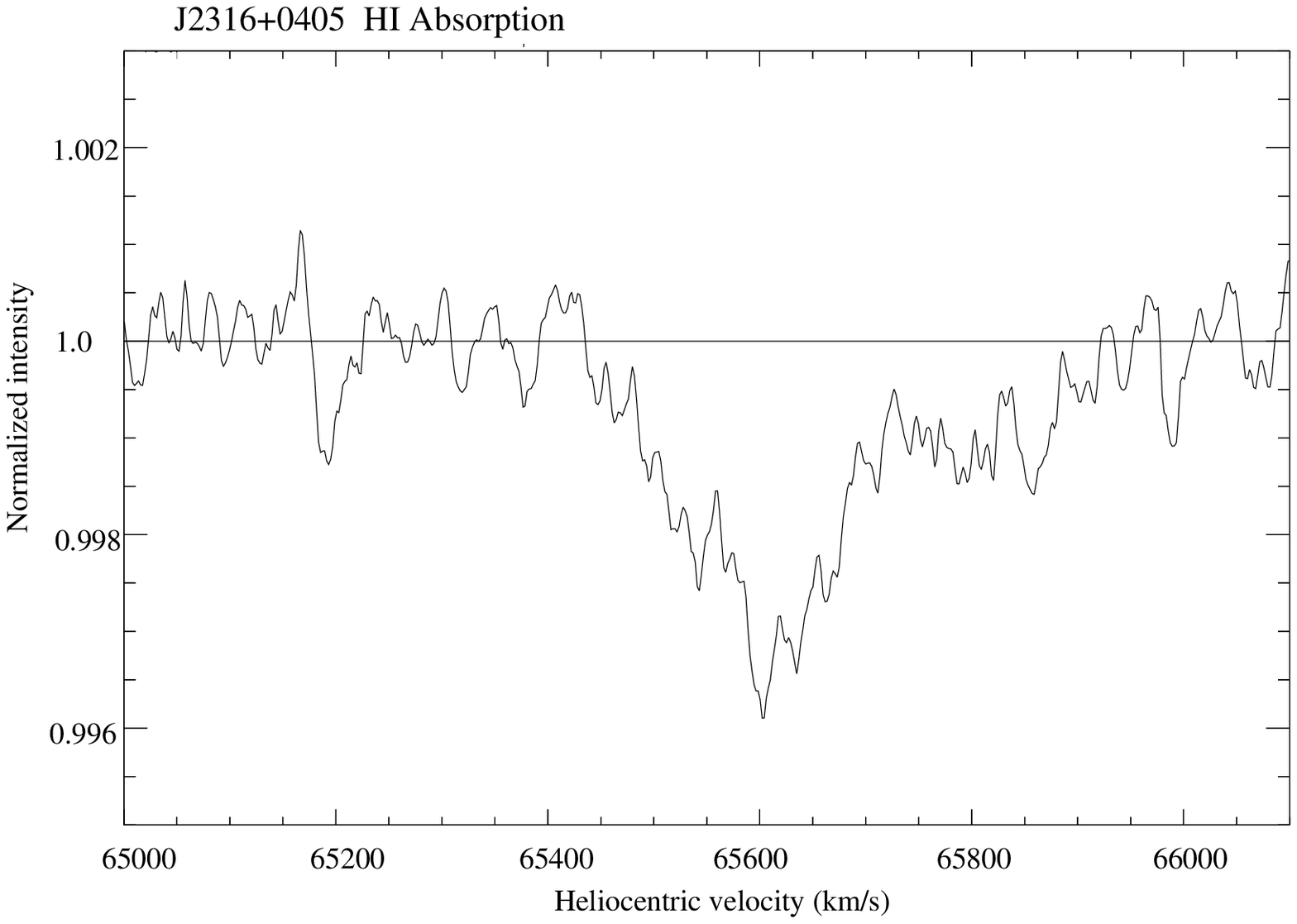,height=7.3cm,width=9.0cm,angle=0}
%\hspace{-4.0cm}
\psfig{figure=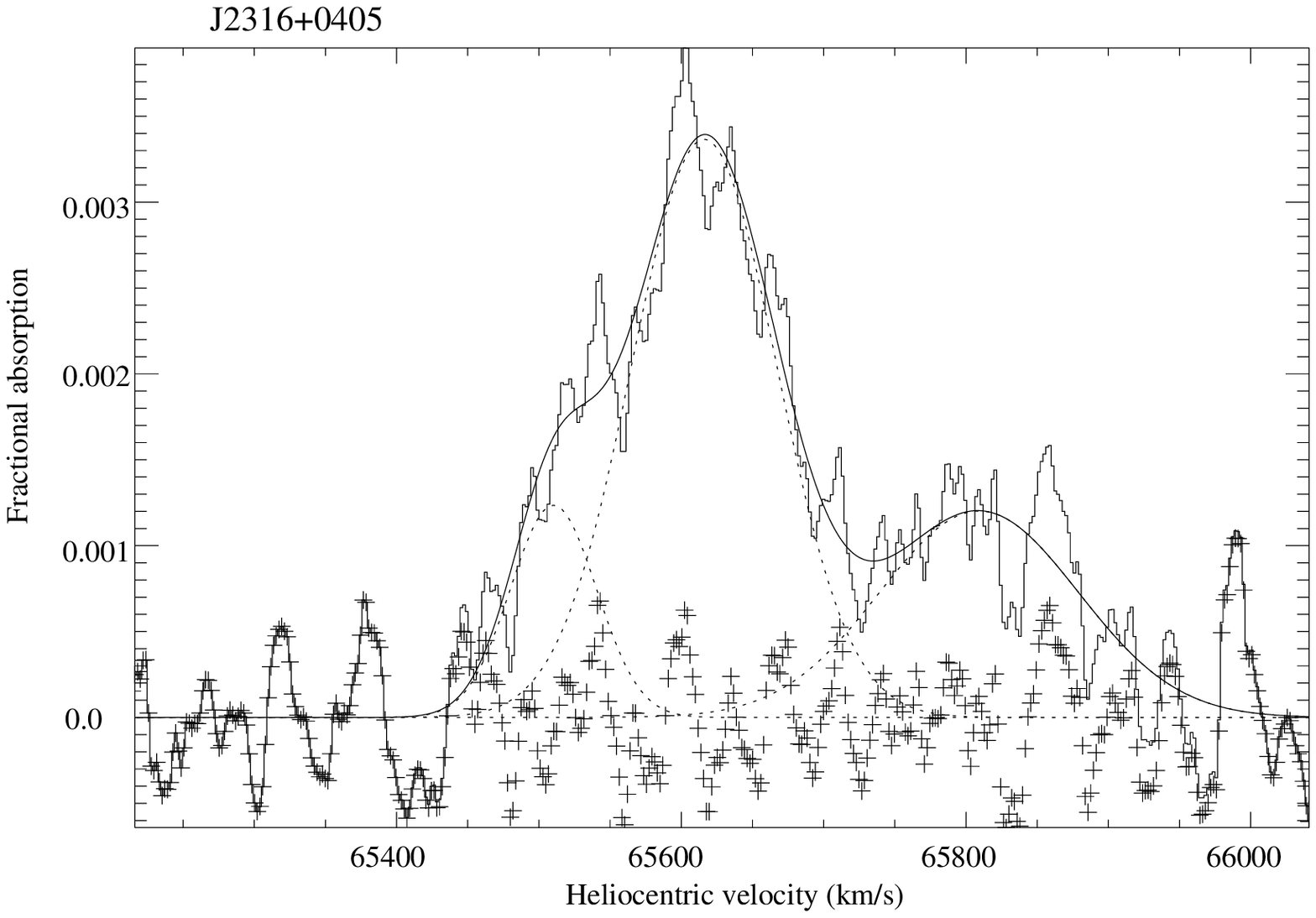,height=7.3cm,width=9.0cm,angle=0}
}}
%\vspace{-2.0cm}
\caption[]{Top: The H{\sc i} absorption spectrum for J2316+0405 (3C459).
The velocity resolution is 17.2 km s$^{-1}$.  Bottom: The same spectrum
overlaid by the three Gaussians fitted to it. The line types are same
as in Fig.~\ref{J0119+3210gfits}.
}
\label{az13}
\end{figure}

%%%%%%
\noindent
{\bf J2250+1419 (4C+14.82):}
This CSS quasar has an angular size of 0$^{\prime\prime}\!$.2 (Spencer et al. 1989).
It was observed with the spectra centered on the then-best redshift, $z = 0.237$. A value of $z = 0.23478(8)$
has recently been published by Eracleous \& Halpern (2004). However, this update is small, and any absorptions
near the revised redshift would still lie well within the present spectra. In practice, 4C+14.82 has no detected
line of either H{\sc i} or OH.  The N(H{\sc i}) limit obtained here is a significant improvement over
the spectrum of Vermeulen et al. (2003).

%%%%%%
\begin{table}
\caption{Multiple-Gaussian fit to the H{\sc i} absorption spectrum of
J2316+0405 (3C459).}
\begin{center}
\vspace{5mm}
\begin{tabular}{|c|l|l|c|c|}
\hline
Id. & v$_{\rm{hel}}$ & FWHM & Frac.  abs. & N(H{\sc i})  \\
no. & km s$^{-1}$ & km s$^{-1}$ & & 10$^{20}$\,cm$^{-2}$ \\
\hline
1 & 65510(27) & 71(34) & 0.0012(0.0001) & 0.17(0.08) \\
2 & 65616(6) & 121(11) & 0.0034(0.0001) & 0.80(0.08) \\
3 & 65809(31) & 164(63) & 0.0012(0.0001) & 0.38(0.15) \\
\hline
\end{tabular}
\end{center}
\label{2314fit}
\end{table}

%%%%%%%
\noindent
{\bf J2316+0405 (3C459):}
This highly-asymmetric radio galaxy with a steep-spectrum core is associated with
a galaxy that exhibits tidal tails and has a high infrared luminosity (Thomasson, Saikia \& Muxlow 2003).
H{\sc i} absorption has been reported for 3C459 by Morganti et al. (2001) and Vermeulen et al. (2003).
The Arecibo H{\sc i} observations confirm the presence of this line, (see Fig.~\ref{az13}: top).  Three
Gaussian components fit the line well, the best-fit parameters being listed in Table~\ref{2314fit}, and shown
in Fig.~\ref{az13}:\,bottom.

%%%
A recently published redshift for this object by Eracleous \& Halpern (2004) gives $z$ = 0.2201 $\pm$ 0.0002
(v $= 65984\pm60$ km s$^{-1}$) from the average of 4 low-ionization lines.  All three fitted H{\sc i}
components are significantly blue-shifted relative to this, consistent with outflow from the galactic nucleus.

%%%%
The OH spectrum for J2316+0405 reveals neither emission nor absorption.  Like J1347+1217, this source was also
in the Arecibo sample of Darling \& Giovanelli (2002), but again no meaningful limit to its OH
emission was set by them due to the effects of standing waves.

%%%%%%%%%%%%%%%%%%%%%%%%%%%%%%%%%%%
\section{Discussion and analyses of the `full sample'}
%%%%%
Of 27 radio sources observed with the Arecibo telescope and GMRT, we detect associated H{\sc i} absorption
towards 6  and H{\sc i} emission for one other. Individual HI emission- and absorption-line systems have been
discussed in detail in Section~\ref{source_notes}.  The source sample we have used contains objects ranging
from the very compact GPS, CSS and CFS sources (sub-kpc to a few kpc) to large radio sources (a few
hundred kpc). Hence, these observations allow us to use radio continuum emission spread over a variety of
scales as background sources to search for associated neutral gas.
In this Section, we combine our results with those of similar H{\sc i} searches and discuss the trends in
gas properties with respect to radio-source characteristics and redshift.

\subsection{The `full sample'}
In order to have a large enough
sample to investigate these trends, we consider the associated H{\sc i}-absorption searches reported
by van Gorkom et al. (1989), Peck et al. (2000), Morganti et al. (2001), Pihlstr\"om (2001),
Vermeulen et al. (2003) and Pihlstr\"om et al. (2003).  These used a variety of instruments and set-ups,
each having different bandwidths, spectral and spatial resolutions.  In the pioneering study of
van Gorkom et al. (1989) a total bandwidth of 6.25 MHz ($\sim$1300 km s$^{-1}$) was used.
It detected H{\sc i} absorption in 4 of 29 galaxies, all 4 being redshifted relative to the systemic velocity.
Vermeulen et al. (2003)  used a bandwidth of 10 MHz, which is amongst the widest, to search for H{\sc i}
absorption towards 57 radio sources.  They detected absorption towards 19 of these  and found a variety of
line profiles from Gaussians with a FWHM of $\lapp$10 \kms ~to irregular and multi-peaked profiles which
sometimes span a few hundred km s$^{-1}$.  However, the typical value is $\sim$150 \kms. Further, the
velocities of the principal (strongest) absorption component with respect to the systemic velocity ranges
from $-$1420 to +318 \kms, with 16 of 19 ($\sim$85\%) lying within the velocity range of $\pm$500 \kms.
In this phase of our study of H{\sc i} absorption towards radio sources, the velocity coverage for the Arecibo
and GMRT observations were typically $\sim$2500 and 1000 km s$^{-1}$ respectively. The GMRT observations would
hence not detect the highest-velocity components of Vermeulen et al., should these be present.

\begin{table*}
\caption{Sample of radio sources from literature forming the `full sample'.}
\begin{center}
\begin{tabular}{|l|l|c|l|l|l|l|l|l|r|r|l|}
\hline
Source & Alt. & Opt. & Redshift & P$_{\rm{5GHz}}$        & LAS                 &  LLS   & Ref. & Radio & N(H{\sc i}) & V$_{shift}$ & Ref. \\
name   & name & ID   &          & 10$^{25}$         &                     &        &      & class & 10$^{20}$ \\
        &      &      &          &   W/Hz            &  $^{\prime\prime}$  &  kpc   &      &       & cm$^{-2}$  &  km s$^{-1}$ & \\
(1)    & (2)  & (3)  &  (4)     & (5)           & (6)                 &  (7)   & (8)  & (9)   & (10) & (11) &  (12)   \\
\hline
J0025$-$2602 & OB-238 & G & 0.322  &  112           & 0.65        & 3.02   &1     & CSS   & 2.42   &$-$30 & V \\
J0111+3906   & OC314  & G & 0.669  &  99            & 0.005       & 0.036  &2,3   & GPS   & 79.8   &   0  & C1\\
              &        &   &        &               &             &        &4,5   &      &        &      &   \\
J0141+1353   & 3C49   & G & 0.621  &  134           & 0.99        & 6.71   &6,7   & CSS   & 1.17   &$-$185& V \\
J0157$-$1043 & OC-192 & Q & 0.616  &  121           & 5.0         & 33.8   &8     & LRG   &$<$1.00 &      & V \\
J0201$-$1132 & 3C57   & Q & 0.669  &  219           & 1.7         & 11.9   &8     & CSS   &$<$0.68 &      & V \\
J0224+2750   & 3C67   & G & 0.310  &  25.8          & 2.5         & 11.3   &7,9   & CSS   &$<$1.45 &      & V \\
J0348+3353   & 3C93.1 & G & 0.243  &  13.6          & 0.56        & 2.14   &10    & CSS   &$<$1.38 &      & V \\
J0401+0036   & 3C99   & G & 0.426  &  33.8          & 4.2         & 23.3   &11    & LRG   &$<$1.13 &      & V \\
J0410+7656   & 4C76.03& G & 0.599  &  336           & 0.14        & 0.93   &7,10  & CSS   & 2.66   &   315& V \\
J0431+2037   & OF247  & G & 0.219  &  32.8          & 0.29        & 1.02   &10,12 & GPS   & 3.66   &   318& V \\
J0503+0203   & OG003  & Q & 0.585  &  196           & 0.011       & 0.073  &13    & GPS   & 6.65   &   43 & C1\\
J0521+1638   & 3C138  & Q & 0.759  &  78.6          & 0.68        & 4.98   &6,7,14& CSS   &$<$0.48 &      & V \\
J0522$-$3627 & MRC & G & 0.0553 & 5.6               & 8.3         & 9.1    &15    & CSS   &$<$0.39 &      & M \\
J0542+4951   & 3C147  & Q & 0.545  &  841           & 0.94        & 5.98   &16    & CSS   &$<$0.30 &      & V \\
J0556$-$0241 &        & G & 0.235  &  4.6           & 0.02        & 0.09   &17    & GPS   &$<$6.27 &      & V \\
J0609+4804   & 3C153  & G & 0.277  &  34.7          & 5.4         & 22.4   &11    & LRG   &$<$0.65 &      & V \\
J0627$-$3529 & MRC & G & 0.0546 & 1.5               & 0.21        & 0.23   &18    & CFS   &$<$1.35 &      & M \\
J0741+3112   & OI363  & Q & 0.635  &  237           & 0.006       & 0.042  &13,19 & GPS   &$<$0.99 &      & V \\
J0815$-$0308 & 3C196.1& G & 0.198  &  3.0           & 3.0         & 9.7    &11    & CSS   &$<$1.53 &      & V \\
J0834+5534   & 4C55.16& G & 0.242  &  85            & 11.0        & 41.4   &20    & LRG   & 1.14   &$-$399& V \\
J0901+2901   & 3C213.1& G & 0.194  &  7.4           & 5.7         & 18.2   &21    & LRG   & 0.12   &$-$14 & V \\
J0909+4253   & 3C216  & Q & 0.670  &  301           & 9.2         & 64.5   &22    & LRG   & 1.30   &   102& V \\
J0927+3902   & 4C39.25& Q & 0.695  &  654           & 3.1         & 22.1   &23    & LRG   &$<$0.99 &      & V \\
J0939+8315   & 3C220.3& G & 0.685  &  145           & 7.8         & 55.1   &24    & LRG   &$<$0.49 &      & V \\
J0943$-$0819 &        & G & 0.228  &  15.5          & 0.020       & 0.072  &25    & GPS   &$<$1.27 &      & V \\
J0954+7435   &        & G & 0.695  &  462           & 0.021       & 0.150  &26,27 & GPS   &$<$2.51 &      & V \\
J1035+5628   & OL553  & G & 0.459  &  72.0          & 0.034       & 0.020  &28    & GPS   &$<$1.38 &      & V \\
J1104+3812   & Mrk421 & G & 0.0300 &  0.14          & 0.075       & 0.044  &29    & CFS   &$<$1.16 &      & G \\
J1120+1420   & 4C14.41& G & 0.362  &  37.7          & 0.084       & 0.420  &30    & GPS   &$<$0.61 &      & V \\
J1159+2914   & 4C29.45& Q & 0.729  &  241           & 5.4         & 39.1   &31    & LRG   &$<$1.91 &      & V \\
J1206+6413   & 3C268.3& G & 0.371  &  55.0          & 1.3         & 6.6    &7,16  & CSS   & 2.07   &   258& V \\
J1220+2916   & NGC4278& G & 0.0022 & 0.0003         & 0.029       & 0.001  &32    & CFS   &$<$2.12 &      & G \\
J1252+5634   & 3C277.1& Q & 0.321  &  26.8          & 1.7         & 7.7    &7,16  & CSS   &$<$0.72 &      & V \\
J1308$-$0950 & OP-010 & G & 0.464  &  166           & 0.60        & 3.47   &1     & CSS   &$<$1.35 &      & V \\
J1313+5458   &        & Q & 0.613  &  72.0          & 0.037       & 0.250  &26    & GPS   &$<$1.80 &      & V \\
J1326+3154   & 4C32.44& G & 0.370  &  88.7          & 0.056       & 0.285  &33,34 & GPS   & 0.75   &$-$471& V \\
J1356+0515   & NGC5363& G & 0.004  & 0.0004         & $<$1.0      & $<$0.08&35    & CFS   &10.3    &   96 & G \\
J1357+4354   &        & G & 0.646  &  58.6          & 0.017       & 0.117  &36    & GPS   & 35.42  &$-$165& V \\
J1400+6210   & 4C62.22& G & 0.431  &  108           & 0.065       & 0.362  &10    & GPS   & 1.99   &$-$258& V \\
J1407$-$2701 & IC4374 & G & 0.022  & 0.056          & 0.009       & 0.004  &25    & CFS   &$<$3.28 &      & G \\
J1415+1320   & OQ122  &G/Q& 0.247  &  12.9          & 0.114       & 0.437  &27,37 & CSS   & 10.60  &   0  & C2\\
J1421+4144   & 3C299  & G & 0.367  &  39.4          & 11.2        & 56.7   &21    & LRG   &$<$0.70 &      & V \\
J1443+7707   & 3C303.1& G & 0.267  &  10.1          & 1.8         & 7.3    &16    & CSS   &$<$1.50 &      & V \\
J1540+1447   & 4C14.60& Q & 0.605  &  120           & 4.0         & 26.8   &38    & LRG   &$<$0.62 &      & V \\
J1546+0026   &        & G & 0.550  &  104           & 0.010       & 0.064  &27    & GPS   &$<$1.05 &      & V \\
J1556$-$7914 & MRC & G & 0.1501 & 24.8              & 0.120       & 0.300  &39    & CFS   &3.82    &{\it a}&M \\
J1642+6856   & 4C69.21& Q & 0.751  &  236           & 8.3         & 61.0   &40    & LRG   &$<$1.34 &      & V \\
J1653+3945   & Mrk501 & G & 0.034  & 0.35           & 0.025       & 0.017  &29    & CFS   &$<$4.44 &      & G \\
J1658+0741   & OS092  & Q & 0.621  &  130           & 7.1         & 48.2   &40    & LRG   &$<$1.56 &      & V \\
J1815+6127   &        & Q & 0.601  &  54.2          & 0.010       & 0.068  &26    & GPS   & 4.61   &$-$1258&V \\
J1816+3457   &        & G & 0.245  &  5.6           & 0.039       & 0.149  &27    & CSS   & 5.40   &$-$184& P1\\
J1819$-$6345 & MRC & G & 0.0645 & 4.2               & 0.33        & 0.400  &1,42  & CSS   &21.2    &$-$160& M \\
J1821+3942   & 4C39.56& G & 0.798  &  319           & 0.47        & 3.53   &10,12 & CSS   & 1.75   &$-$806& V \\
J1823+7938   &        & G & 0.224  &  6.3           & 0.015       & 0.054  &26    & GPS   &$<$26.7 &      & V \\
J1829+4844   & 3C380  & Q & 0.692  &  912           & 1.3         & 9.0    &16    & CSS   &$<$0.19 &      & V \\
J1831+2907   & 4C29.56& G & 0.842  &  357           & 2.6         & 19.9   &43    & LRG   &$<$1.70 &      & V \\
J1845+3541   & OU373  & G & 0.764  &  126           & 0.010       & 0.071  &44    & GPS   &$<$11.0 &      & V \\
J1944+5448   & OV573  & G & 0.263  &  17.3          & 0.040       & 0.161  &12,44 & GPS   & 5.21   &$-$1420&V \\
J1945+7055   &        & G & 0.101  &  1.5           & 0.035       & 0.064  &45    & GPS   & 30.0   &$-$172& P2\\
\hline
\end{tabular}
\end{center}
\label{fsample}
\end{table*}
\begin{table*}
\contcaption{}
\begin{center}
\begin{tabular}{|l|l|c|l|l|l|l|l|l|r|r|l|}
\hline
Source & Alt. & Opt. & Redshift & P$_{\rm{5GHz}}$        & LAS                 &  LLS   & Ref. & Radio & N(H{\sc i}) & V$_{shift}$ & Ref. \\
name   & name & ID   &          & $\times$10$^{25}$ &                     &        &      & class & $\times$10$^{20}$ \\
        &      &      &          &   W/Hz            &  $^{\prime\prime}$  &  kpc   &      &       &   cm$^{-2}$  &   km s$^{-1}$  & \\
(1)    & (2)  & (3)  &  (4)     & (5)           & (6)                 &  (7)   & (8)  & (9)   & (10) & (11) &  (12)   \\
\hline
J2022+6136   & OW637  & Q & 0.227  &  32.4          & 0.010       & 0.036  &28    & GPS   &$<$0.38 &      & V \\
J2052+3635   &        & G & 0.355  &  117           & 0.060       & 0.297  &46    & GPS   & 7.69   &$-$95 & V \\
J2137$-$2042 & OX-258 & G & 0.635  &  223           & 0.200       & 1.370  &1,47  & CSS   &$<$1.18 &      & V \\
J2151+0552   &        & G & 0.74   &  112           & 0.004       & 0.027  &48    & GPS   &$<$16.9 &      & C1\\
J2202+4216   & DA571  & Q & 0.069  & 3.8            & 0.010       & 0.013  &2     & CFS   &$<$1.74 &      & G \\
J2209$-$2331 & MRC    & G & 0.087  & 1.7            &$<$2.0       &$<$3.2  &49    & CSS   &$<$1.35 &      & G \\
J2255+1313   & 3C455  & Q & 0.543  &  102           & 2.6         & 16.5   &21    & LRG   & 0.45   &   30 & V \\
J2257$-$3627 & MRC    & G & 0.006  & 0.009          & 0.030       & 0.004  &47    & CFS   &$<$1.74 &      & G \\
J2321+2346   & 3C460  & G & 0.268  &  9.4           & 5.8         & 23.7   &41    & LRG   &$<$2.04 &      & V \\
J2333$-$2343 & OZ-252 & G & 0.048  & 0.46           & 0.037       & 0.034  &28    & CFS   &$<$1.54 &      & G \\
J2344+8226   &        & Q & 0.735  &  292           & 0.267       & 1.940  &10,12 & GPS   &$<$0.75 &      & V \\
J2355+4950   & OZ488  & G & 0.238  &  22.3          & 0.069       & 0.256  &50,51 & GPS   & 3.01   &$-$12 & V \\
\hline
\end{tabular}
\end{center}
\begin{flushleft}
Col. 1: source name; col. 2: alternative name; col. 3: optical identification;
col. 4: redshift; col. 5: 5-GHz luminosity in the rest frame of the source;
cols. 6 \& 7: largest projected angular (LAS) and linear (LLS) size
in arcsec and kpc respectively, as measured from the outermost radio peaks; col. 8: references for the radio structure;
col. 9:  radio class; col. 10: H{\sc i} column density or a 3$\sigma$ upper limit to it;
col. 11: the shift of the primary H{\sc i} component relative to the systemic velocity as measured from the optical
emission lines, with a negative sign indicating a blue-shift, and column 12: references for the H{\sc i} observations.\\
References for the structural information and LLS: 1: Tzioumis et al. (2002); 2: Pearson \& Readhead (1988);
3: Owsianik, Conway \& Polatidis (1998); 4: Zensus et al. (2002); 5: Baum et al. (1990); 6: Fanti et al. (1989);
7: Saikia et al. (1995); 8: Reid, Kronberg \& Perley (1999); 9: Sanghera et al. (1995); 10: Dallacasa et al. (1995);
11: Neff, Roberts \& Hutchings (1995); 12: Saikia et al. (2001); 13: Stanghellini et al. (2001);
14: Akujor et al. (1991); 15: Keel (1986); 16: L\"udke et al. (1998); 17: Petrov et al. (2005); 18: Venturi et al. (2000);
19: Saikia \& Kulkarni (1998); 20: Whyborn et al. (1985); 21: Akujor \& Garrington (1995); 22: van Breugel et al. (1992);
23: Browne et al. (1982); 24: Jenkins, Pooley \& Riley (1977); 25: Beasley et al. (2002);
26: Taylor et al. (1994);
27: Peck et al. (2000); 28: Fomalont et al. (2000); 29: Giovannini, Feretti \& Venturi (1999);
30: Bondi, Garrett \& Gurvits (1998); 31: Antonucci \& Ulvestad (1985); 32: Giovannini et al. (2001);
33: Mutel et al. (1981); 34: Fey, Clegg \& Fomalont (1996); 35: van Gorkom et al. (1989);
36: Taylor et al. (1996);
37: Perlman et al. (1996); 38: Ulvestad, Johnston \& Weiler (1983);
39: Morganti et al. (2001); 40: Murphy, Browne \& Perley (1993); 41: Giovannini et al. (1988);
42: Ojha et al. (2004); 43: Spencer et al. (1989); 44: Xu et al. (1995); 45: Taylor \& Vermeulen (1997);
46: Philips \& Mutel (1981); 47: Fomalont et al. (2003); 48: Stanghellini, O'Dea \& Murphy (1999);
49: Kapahi et al. (1998a); 50: Polatidis et al. (1995); 51: Taylor et al. (2000). \\
References for H{\sc i} observations:  C1: Carilli et al. (1998); C2: Carilli et al. (1992);
M: Morganti et al. (2001); G: van Gorkom et al. (1989); M: Morganti et al. (2001);
P1: Peck et al. (2000); P2: Peck, Taylor \& Conway (1999); V: Vermeulen et al. (2003) \\
{\it a} Errors in the systemic velocity are too large.
\end{flushleft}
%
%\label{fsample}
\end{table*}
In the samples considered here, the spatial resolution was usually insufficient to resolve compact sources,
while larger sources are often resolved. In order to have an average value of the H{\sc i} optical depth,
or an upper limit over the scale length of the radio source, we have considered only sources for which most
of the radio continuum emission lies well within the resolution element of the telescopes. In some sources,
H{\sc i} absorption may be detected against a single component, such as the radio cores in PKS1318$-$43 and
3C353 (e.g. Morganti et al. 2001). We will consider such cases in a later paper, along with our H{\sc i}
absorption observations towards the cores and hotspots of radio sources.

\begin{figure}
\centerline{\vbox{
\psfig{figure=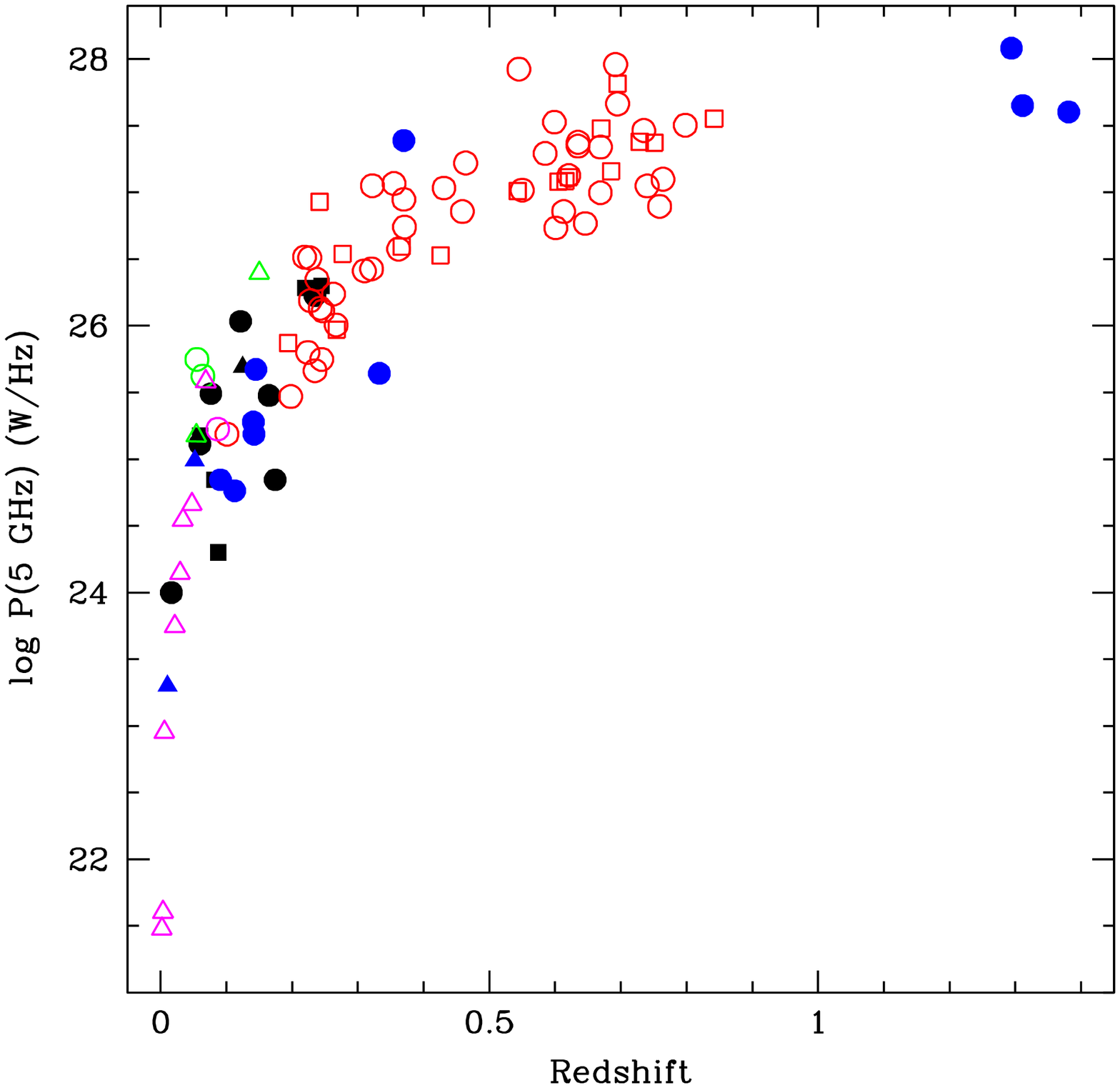,height=9.3cm,width=9.0cm,angle=0}
}}
\caption[]{Luminosity of sample sources at 5 GHz versus redshift.  The solid symbols are sources from
our observations, while open symbols are data from the literature.  The circles denote GPS/CSS, triangles CFS,
and squares LRG sources.  For our observations, black represents Arecibo sources and blue the GMRT's.
Red corresponds to data from Vermeulen et al. (2003) and Pihlstr\"om et al. (2003), green to
Morganti et al. (2001) and magenta to van Gorkom et al. (1989).
}
\label{powervsz}
\end{figure}

With the above selection criterion, we obtain a sample of 96 radio sources, called the `full sample',
which is a heterogeneous superset of sources with available H{\sc i} information.
This sample is the combination of Tables~\ref{sample} and \ref{fsample}, with the omission of
J1643+1715 for which no H{\sc i} data exists.  Data for J0901+2901 was taken from
Vermeulen et al. (2003), rather than from our own shallower observations of the source.
The sample includes 27 GPS, 35 CSS, 13 CFS and
21 LRG sources (see Table~\ref{fullsample}).  Amongst the GPS and CSS objects, 17 are associated with
quasars, and 44 with galaxies. All the CFS sources are associated with quasars or BL Lac objects,
while the LRGs are associated with galaxies.  The sample spans a wide range of $\sim$6 orders of magnitude
in luminosity at 5 GHz and $z\lapp$1.4. However, since most sources have been selected from strong flux-density
limited samples, luminosity and redshift are strongly correlated (see Fig.~\ref{powervsz}).

%
%%%%%%%%%%%%%%%%%%%%

%%%%%
\subsection{H{\sc i} column density versus linear size}
\label{NHI-z}
\begin{figure}
\centerline{\vbox{
\psfig{figure=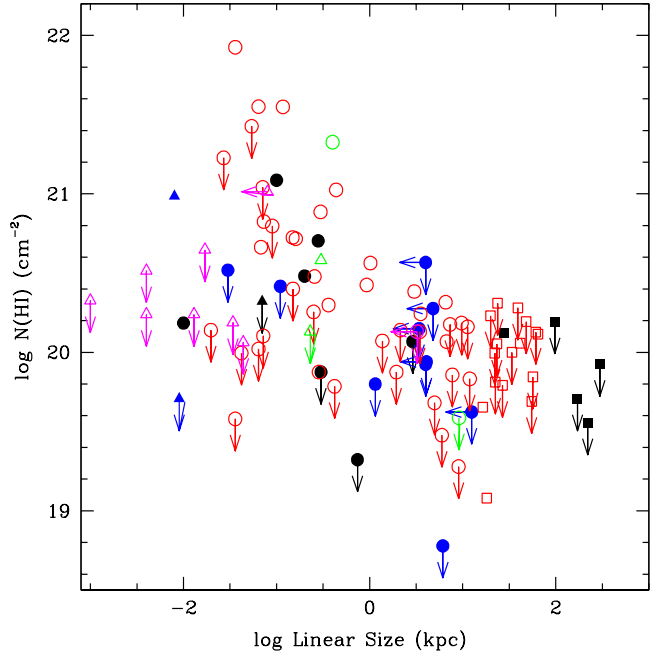,height=9.3cm,width=9.0cm,angle=0}
}}
\caption[]{H{\sc i} column density as a function of projected linear source size.  Arrows mark the 3$\sigma$
upper limit to the column density or to  the linear size.  See the caption of Fig.~\ref{powervsz} for the meaning
of the colours and symbols.
}
\label{nhivssize}
\end{figure}

For the `full sample', Fig.~\ref{nhivssize} shows the 21-cm H{\sc i} column density against projected linear
size measured from the outermost peaks of radio emission.  T$_{\rm s} =$100 K and full coverage of the background source
have been assumed to estimate the column density using eqn.~\ref{eqcol}.  The upper limits on N(H{\sc i}) 
for non-detections
are calculated from the 3$\sigma$ limit on the peak optical depth as estimated from the spectrum or given in the
literature, and a velocity width of 100~km~s$^{-1}$.
Fig.~\ref{nhivssize} shows that the incidence of H{\sc i} absorption is much higher for the compact sources.
In particular, for sources with sizes less than 15 kpc, H{\sc i} absorption is detected towards 26 out of 75
($\sim$35\%),  while only 5 out 21 ($\sim$25\%) LRG sources exhibit H{\sc i} absorption.  The detection rate
($\sim$45\%) is maximum for the most compact class, the GPS sources (see Table~\ref{fullsample}).

The `full sample' shows an anticorrelation between N(H{\sc i}) and source size 
(Fig.~\ref{nhivssize}), which was first noted by Pihlstr\"om et al. (2003) for the GPS and CSS sources.
However, when interpreting this relationship, we need to bear in mind the 
uncertainties in the values of spin temperature and covering factor for the different objects. 
The small sources are capable of probing gas closer to the AGN so that T${\rm _s}$ could be greater than 100 K. 
If this were so, it would only increase our estimate of the column density for the smaller sources 
thereby increasing
the difference in H{\sc i} column density between the smaller and the larger objects. The other factor
influencing this relationship is the covering factor which could in principle increase the estimates of
column density for the larger objects. This depends on the size of the absorber relative to the radio
source and can be probed by high-resolution spectroscopic observations. 

It is relevant to note that the above anti-correlation is consistent with the optical spectroscopic observations 
of van Ojik et al. (1997), who interpreted deep troughs in the Ly$\alpha$ emission profiles of 18 high-redshift
($z~>$ 2) radio galaxies (HZRGs) to be absorption due to H{\sc i} with column densities in the range
10$^{18}-$10$^{19.5}$ cm$^{-2}$.  Interestingly, 9 out of the 10 HZRGs with sizes $<$50 kpc have associated
H{\sc i} absorption compared with 2 of the 8 larger ($>$50 kpc) ones.

%%%%%%%

%%%%%

%%%%%%
\begin{figure}
\centerline{\vbox{
\psfig{figure=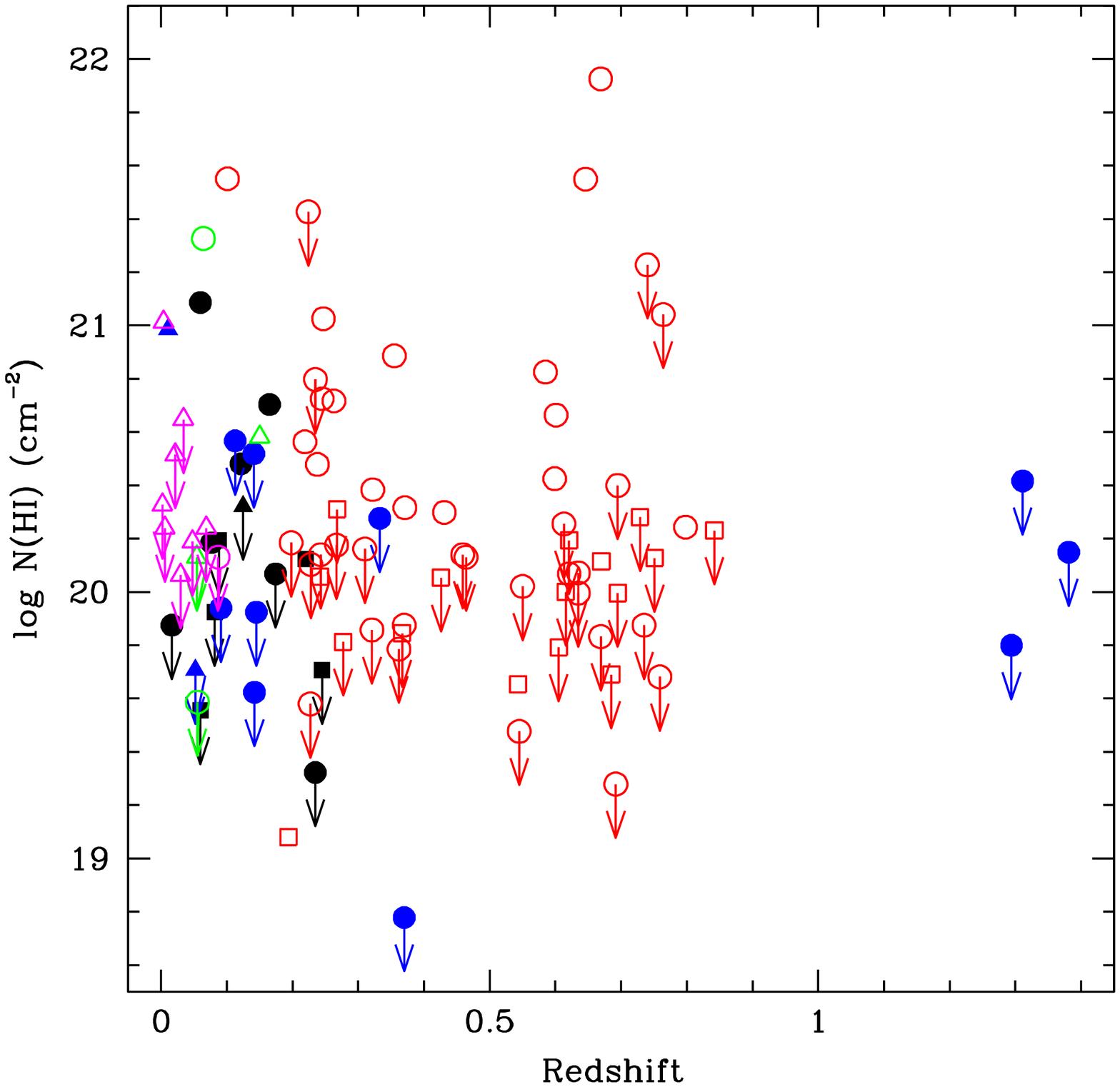,height=9.3cm,width=9.0cm,angle=0}
}}
\caption[]{H{\sc i} column density versus redshift.  See the caption of Fig.~\ref{powervsz} for the meaning
of the colours and symbols.}
\label{nhivsz}
\end{figure}

%%%%%%

\subsection{H{\sc i} column density versus redshift}
\begin{figure}
\centerline{\vbox{
\psfig{figure=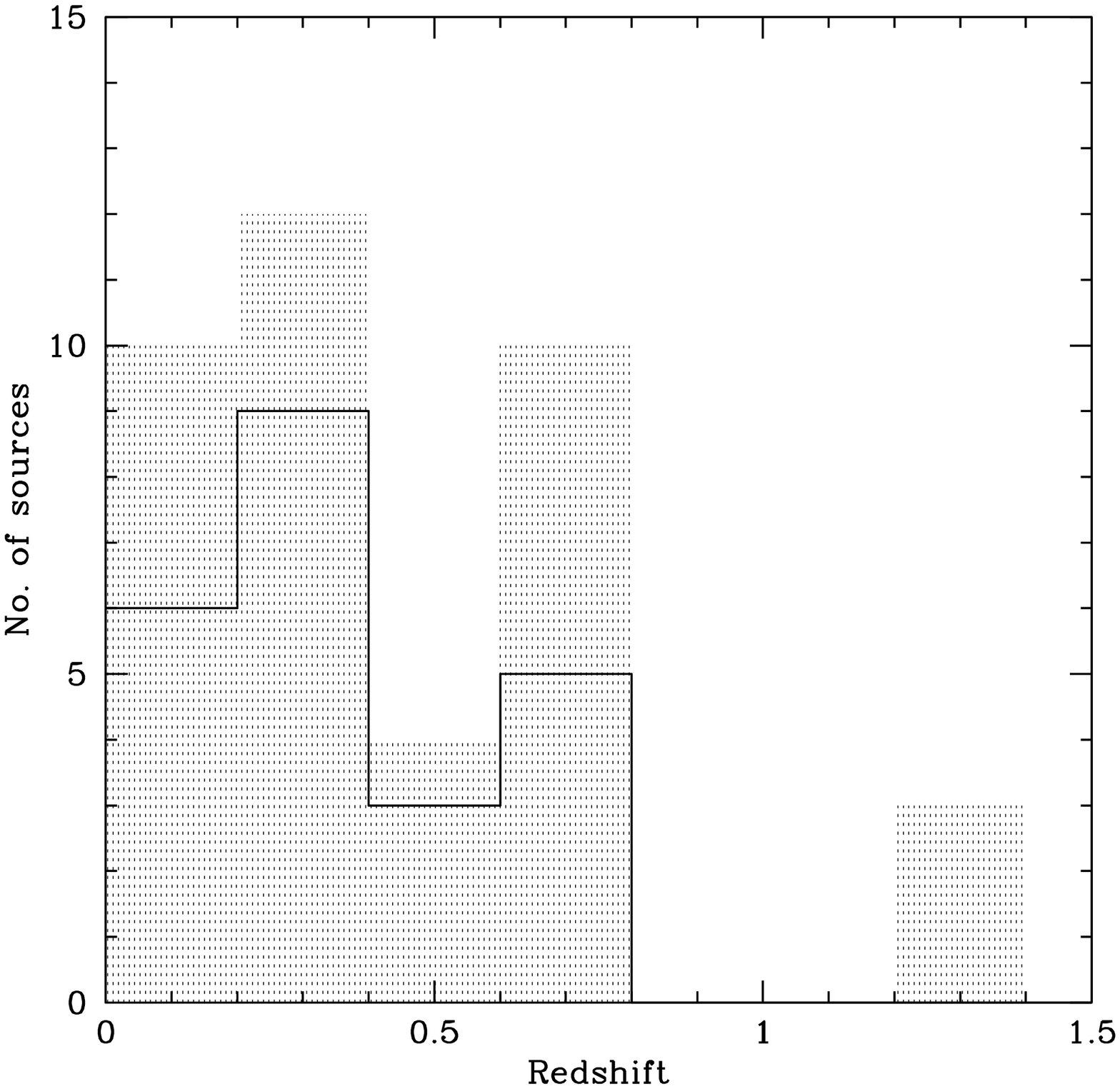,height=9.3cm,width=9.0cm,angle=0}
}}
\caption[]{The redshift distribution of GPS and CSS sources with detections (solid lines) and
non-detections (shaded).}
\label{cssgpshist}
\end{figure}

The HST studies of distant galaxies (z$~\lapp$1; e.g. Abraham et al. 1996; Brinchmann et al. 1998) show
remarkable morphological evolution and suggest larger incidences of mergers and interactions in the past.
Such mergers and interactions could trigger gas inflows towards the central regions of galaxies leading to
the formation of circumnuclear starbursts, tori and the fuelling of nuclear black holes (Sanders et al. 1988;
Hopkins et al. 2005).  It is thus pertinent to enquire whether the environments of radio sources as probed
by H{\sc i} absorption bear any signature of evolution with redshift.
In Fig.~\ref{nhivsz}, we plot N(H{\sc i}) against redshift for the `full sample'.  Clearly, no trend exists
for N(H{\sc i}) to vary with redshift out to z$\sim$1.
However, it is possible that any such dependence is not apparent for the `full sample' as it contains sources
with luminosities and projected linear sizes spanning a range of $\sim$6 orders of magnitude in luminosity
and 4$-$5 orders of magnitude in linear size.  The parameter space can be constrained to some extent by considering
the sub-sample consisting of just GPS and CSS sources.  This sub-sample contains 62 sources with projected
linear size $\le$ 15 kpc, $\sim$85\% of which have luminosities spanning $\sim$2 orders of magnitude.
In Fig.~\ref{cssgpshist}, we plot the redshift distributions of these GPS and CSS sources separately for
H{\sc i} detections and non-detections.  There is again no significant difference between the two redshift
distributions, suggesting that the environments of radio sources on GPS and CSS scales as probed by the neutral
gas component are similar at different redshifts.
Interestingly, this is consistent with the polarization asymmetry study of CSS and larger sources
by Saikia \& Gupta (2003), whose sample of sources with z$~\lapp$ 2, showed no significant dependence
of polarization asymmetry on redshift.  A study of the fraction of CSS objects at different redshifts in
well-defined complete samples led R\"ottgering et al. (1996) to suggest that the environments of CSS
objects do not vary significantly with redshift.
%%

%%%%%%%

\subsection{H{\sc i} column density versus luminosity}
%%%%%%
\begin{figure}
\centerline{\vbox{
\psfig{figure=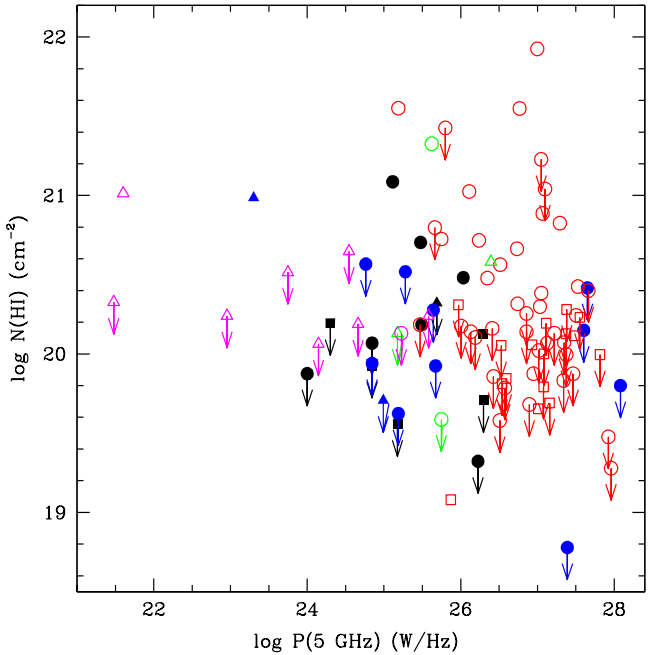,height=9.3cm,width=9.0cm,angle=0}
}}
\caption[]{H{\sc i} column density versus source luminosity at 5 GHz.  See the caption of Fig.~\ref{powervsz} for
the meaning of the colours and symbols.}
\label{nhivspower}
\end{figure}
%%%%%

%%
%%%%%%%%%
%
\begin{table}
\caption{Summary of the `full sample'.}
\begin{center}
\begin{tabular}{lcc}
\hline
  Radio class.& Detections  & Non-detections   \\
\hline
GPS          &   12        &    15            \\
CSS          &   11        &    24            \\
CFS          &   3         &    10            \\
LRG          &   5         &    16            \\
\hline
\end{tabular}
\end{center}
\label{fullsample}
\end{table}
It is widely believed that the energy output from AGN is triggered by the supply of gas to the central engine,
presumably a supermassive black hole (e.g. Rees 1984).  Although there have been a number of studies of gas
kinematics of nearby AGN in order to understand the fuelling of such activity, the dependence of N(H{\sc i}) on
radio luminosity could provide clues towards understanding this phenomenon.
In Fig.~\ref{nhivspower}, we plot N(H{\sc i}) against the radio luminosity at 5 GHz for the `full sample'.
There is no significant dependence of N(H{\sc i}) on radio luminosity. Since luminosity and redshift are strongly
correlated in our sample, and there is no signficant dependence of N(H{\sc i}) on redshift, this is not surprising.
In order to investigate any dependence of N(H{\sc i}) on luminosity, one requires objects with H{\sc i} detections
covering a large range in luminosity but confined to a narrow range in redshift for a given class of objects.
Although the `full sample' contains objects spanning $\sim$4 orders of magnitude in luminosity in the
lowest redshift range, these are a ``mixed bag'' with very few H{\sc i} detections.
To address this issue satisfactorily, one needs to extend the H{\sc i} observations of radio sources,
especially GPS and CSS sources, to lower luminosities of (say) 10$^{24}$ to 10$^{26}$ W Hz$^{-1}$
at 5 GHz in the redshift range of $\sim$0.5 to 1.

\subsection{H{\sc i} absorption and unification scheme}
%%%%%%%
\begin{figure}
\centerline{\vbox{
\psfig{figure=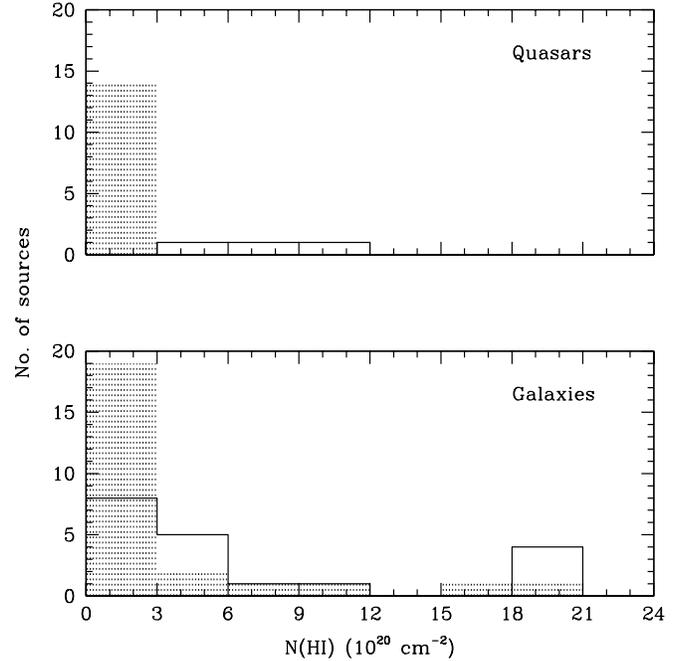,height=9.3cm,width=9.0cm,angle=0}
}}
\caption[]{The H{\sc i} column density distribution of GPS and CSS quasars (top panel) and galaxies (bottom panel).
The detections are marked by solid lines while the shaded regions represent non-detections.  The four detections
and one non-detection for galaxies with log~N(H{\sc i})$>$19.5 have been put in the last bin.
}
\label{cssQGhist}
\end{figure}

%%%%%%
GPS and CSS sources are believed to be young, seen at an early stage of their evolution.  Though it
is possible that a few (e.g. 3C216) have small linear sizes due to projection effects,
most are intrinsically compact (Fanti et al. 1990). An interesting question is whether GPS and CSS objects
are consistent with the unification scheme for radio galaxies and quasars, in which quasars are inclined
at $\lapp$45$^\circ$ to the line of sight while galaxies lie at larger angles (Barthel 1989; Antonucci 1993).
Saikia et al. (1995, 2001) have argued from the prominence of their radio cores and their symmetry parameters
that GPS and CSS objects are also consistent with the unified scheme. If this is so, then one might also
expect a dependence of the detection rate of H{\sc i} absorption and H{\sc i} column density on the optical
identification of the parent optical object and the fraction of emission from the core (Pihlstr\"om et al. 2003),
which is a reasonable statistical indicator of the source-axis orientation to the line of sight
(Orr \& Browne 1982; Kapahi \& Saikia 1982). The exact nature of the variation would depend on the form of the
disk or torus, which is not well constrained.

As a first step, we examine whether the detection rate for the `full sample' is consistent with
the unification scheme.  In Fig.~\ref{cssQGhist}, we plot the N(H{\sc i}) distributions for GPS and CSS
quasars and galaxies.  Clearly, the detection rate ($\sim$40\%) for galaxies is higher than that ($\sim$20\%)
for quasars. Considering just the GPS and CSS objects separately, one finds for both a similar difference in
the detection rates for galaxies and quasars. This is consistent with the unification scheme for radio
galaxies and quasars (e.g. Pihlstr\"om et al. 2003). In addition to the geometries considered by
Pihlstr\"om et al.,  the distribution of H{\sc i} gas clouds is likely to be anisotropic since those close
to the path of the radio jet, or within the ionisation cone, are more likely to be ionised than those
at larger angles to the jet axis (cf. van Ojik et al. 1997). This could also cause a higher
detection rate of H{\sc i} absorption in galaxies relative to quasars in the framework of orientation-based
unification schemes.

Gupta \& Saikia (2006) have recently reported the consistency of H{\sc i} properties with the unification scheme for 
radio galaxies and quasars selected from this `full sample'. 
They have shown that there is a tendency for the detection rate as well as the column density for galaxies 
to increase with core prominence, a statistical indicator of the orientation of the jet axis to the line of 
sight. This can be understood in a scenario where radio sources are larger than the scale of the circumnuclear 
H{\sc i} disc so that the lines of sight to the lobes at very large inclinations do not intersect the disc. 
They also suggest that small linear size, along with intermediate values of core prominence, is a good recipe 
for detecting 21-cm absorption in CSS and GPS objects. 

%%%%%%%%%%%%%%%%%%%%%%%
%%
%%%%%%%
\begin{figure}
\centerline{\vbox{
\psfig{figure=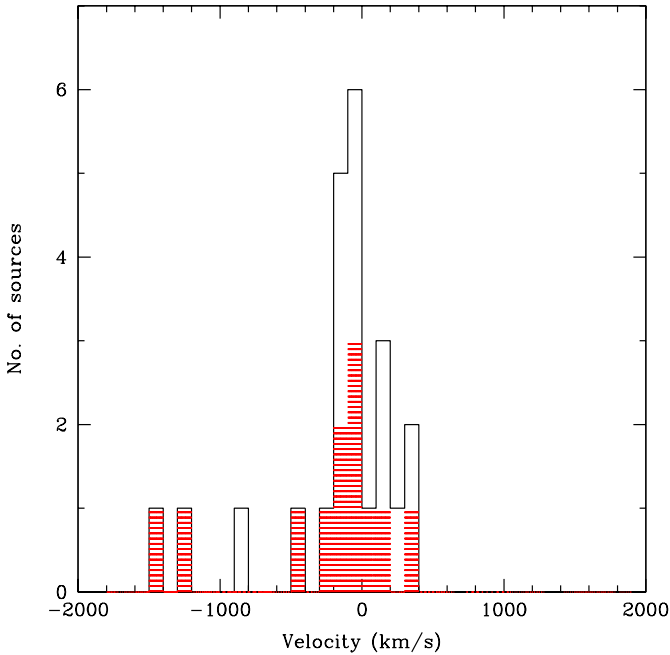,height=9.3cm,width=9.0cm,angle=0}
}}
\caption[]{The distribution of velocities for the principal H{\sc i} absorption components with respect to the
systemic velocity estimated from optical lines.   The solid line is the distribution for both CSS and GPS sources,
while the distribution for GPS sources is shown shaded.
}
\label{velhist}
\end{figure}

\subsection{Blue-shifted absorption lines: Evidence of jet-cloud interaction?}
Although blue-shifted absorption lines could arise due to either halo gas or circumnuclear gas kinematically
affected by nuclear winds and/or radiation pressure, the jet-cloud interaction may also play a significant
role in sub-galactic sized radio sources.
The structure of GPS and CSS sources often
appear to be affected by the ambient gas, suggesting dynamical interaction with the external medium
(Spencer et al. 1991; Schilizzi et al. 2000).  In particular, CSS sources tend to be more asymmetric in
brightness, location and polarization of the outer radio components when compared with the larger
sources (Saikia et al. 1995, 2001; Arshakian \& Longair 2000; Saikia \& Gupta 2003).  They also
tend to exhibit high rotation measures (see O'Dea 1998 for a review).
The properties of emission-line gas in the galaxies associated with these sources are also consistent
with the scenario in which ambient gas has been ionized and accelerated by shocks caused by interaction
with the radio jet (Gelderman \& Whittle 1994; Labiano et al. 2005).
In the archetypal CSS object 3C48, the kinematical properties of the blue-shifted gas seen in emission and
absorption (Chatzichristou et al. 1999;  Gupta et al. 2005) suggest that the complex radio structure of
the source (Wilkinson et al. 1991) could be caused by interaction of the jet with the external gas clouds.
Baker et al. (2002) studied absorption spectra for a sample of quasars from the Molonglo Reference
Catalogue, and find a slight excess of blue-shifted C~{\sc iv} absorption lines in CSS sources as compared with
the larger objects.  van Ojik et al. (1997) also found an excess of blue-shifted Ly$\alpha$ absorption lines
in their sample of HZRGs, and demonstrated strong evidence for interaction of the radio sources with the
external environment.  In principle jet-cloud interaction can give rise to blue-shifted as well as
red-shifted absorption lines.  However, 21-cm absorption line experiments towards a sample of sources which
are randomly oriented towards us could exhibit an excess of blue-shifted lines.
This is  because for the sources oriented significantly away from the sky plane,
a) the approaching jet will be stronger than the receding jet due to
Doppler boosting and b) the absorbing gas towards the approaching hot-spot/lobe is likely to have a larger
covering factor.  These effects possibly play a role in the fact that broad ($\sim$1000 km~s$^{-1}$) 21-cm
absorption lines towards radio sources, which are believed to have been caused by jet-cloud interaction,
are all blue-shifted (Morganti, Tadhunter \& Oosterloo 2005).

In Fig.~\ref{velhist}, we plot the distribution of the velocity shifts of the principal (strongest)
21-cm absorption
component detected towards the CSS and GPS sources in the `full sample'.  Though
hampered by small number statistics,
and uncertainties in systemic velocities (cf. Mirabel 1989; Morganti et al. 2001), the distribution does indicate
that H{\sc i} absorption for compact radio sources tends to be blue-shifted, with 14 out of the 23 principal
absorption-line components displaying such an effect.
We note that the skewness of the distribution comes largely from the principal components detected towards GPS
sources.  Of the 12 principal components detected towards GPS sources 8 have blue-shifted velocities with respect
to the systemic velocity.  Also, 6 of the 9 secondary components listed by Vermeulen et al. (2003) appear
blue shifted.
The higher incidence of blue-shifted features suggests that some of the absorption could be due to gas clouds
accelerated by the radio jets.

%%
%%%%%%%
%%%%%%%%%%%%%%%%%%%%%%%%%%%%%%%%%%%%%%%
\subsection{OH upper limits and their implications}
\label{ohimp}
The parameters of the 9 OH non-detections at Arecibo are given in Table~\ref{obsArOH}.  Upper limits on thermal
OH column densities can be calculated using
\begin{equation}
{\rm N(OH)} = 2.25\times10^{14}\frac{{\rm T_{ex}}\int{\tau_{1667}(v)dv}}{f_c} \, {\rm cm^{-2}   }
\label{eqoh}
\end{equation}
where $\tau_{1667}$(v) is the optical depth in the 1667.3-MHz OH main-line at radial velocity $v$\,km s$^{-1}$
(e.g. Boyce \& Cohen 1994).  Below, a covering factor of $f_c$=1.0 and an excitation temperature of
${\rm T}_{\rm{ex}} =10$\,K are adopted.

Assuming that a velocity equivalent width from the H{\sc i} absorption of
$v_{eq} = \int\!\tau_{H{\sc I}}(v)\,dv / \tau_{p}$
also applies to any OH absorption, we have computed upper limits on the OH column densities for the three sources
for which both an H{\sc i} absorption was detected at Arecibo and an upper limit obtained on OH absorption. An
optical depth limit for OH absorption of {\rm $3\times\sigma_{\rm{FA}}^{\rm{OH}}$} was adopted,
where {\rm $\sigma_{\rm{FA}}^{\rm{OH}}$} is taken from Table~\ref{obsArres}.  This gives column densities of
$<1.1\times10^{14}$\,cm$^{-2}$ for J1124+1919, $<3.3\times10^{14}$\,cm$^{-2}$ for J1347+1217, and
$<10.9\times10^{14}$\,cm$^{-2}$ for J2316+0405. Of course, these upper limits need to be interpreted in the light
of the expected covering factors for potential absorbers obtained from VLBI imaging and the projected
angular sizes for diffuse clouds and giant molecular clouds (GMCs).  However, we note that in our Galaxy, the
column densities of OH in respect of diffuse clouds are of the order of N(OH)$\sim10^{13-14}$\,cm$^{-2}$,
while typical values for GMCs are N(OH)$\sim10^{15-16}$\,cm$^{-2}$.  The above column densities for H{\sc i}
and OH imply an abundance ratio of N(H{\sc i})/N(OH) $\ge4.4\times10^{6}$ for J1124+1919, the source with the
lowest limit on N(OH), given the values adopted for H{\sc i} and OH temperatures, and assuming similar
covering factors.  Other upper limits for N(OH) in Table~\ref{obsArOH} are 3-$\sigma$ values, assuming a velocity
width of 100 km~s$^{-1}$.  It is worth noting for comparison that although there are only a few OH main-line 
absorption systems seen towards extragalactic sources, the typical column density for the detected ones is 
a few times $\sim$10$^{15}$ cm$^{-2}$ (e.g. Darling \& Giovanelli 2002; Kanekar \& Chengalur 2002).

The non-detection of megamaser emission in the 1667-MHz OH main line provides us with upper limits on
(isotropic) luminosities for this line. Adopting exactly the prescription used for their non-detections
by Darling \& Giovanelli (2002), which assumes a line intensity of 1.5\,$\sigma$, a rest-frame velocity width
of 150 kms$^{-1}$, and normalization to H$_{0} = 75$\,km~s$^{-1}$Mpc$^{-1}$, $q = 0$, upper limits have been
computed for the 9 sources for which good OH spectra were obtained.  These luminosity limits,
L$_{\rm{OH}}\!\!\!\!\!\!\!^{^{\rm{max}}}$, are given in Table~\ref{obsArOH} and agree within 5\% with
the estimated luminosities were the same cosmological parameters to be adopted as in the rest of this paper.
Of these, only J1347+1217 and
J2316+0405 were detected by IRAS. From the L$_{\rm{OH}}$ -- L$_{\rm{FIR}}$ relationship of Kandalian (1996),
Darling \& Giovanelli predict luminosity values for these of L$_{\rm{OH}} = 2.36$ and
2.46\,$h^{^{-2}}\!\!\!\!\!_{75}$\,L$_{\odot}$ respectively.  While our luminosity upper limit for J2316+0405
is about a factor of 4 higher than the predicted OH luminosity, that for J1347+1217 is somewhat below the
predicted value. As mentioned in Section 3.3, obtaining a spectrum of increased sensitivity for J1347+1217
would be valuable.

In respect of the non-detection of OH emission for any of the AGNs observed by us, we note that
Darling \& Giovanelli (2000) failed to detect any OH megamasers for a sample of nearby AGNs situated in
quiescent (non-interacting) systems and which were undetected by IRAS.

\begin{table}
\caption{Results of the Arecibo search for associated OH absorption.  }
\begin{center}
\begin{tabular}{|l|c|c|c|r|c|}
\hline
Source &  Date &$\sigma_{\rm{FA}}^{\rm{OH}}$ & $\Delta v$$^{\rm{OH}}$ & N(OH) & log \\
name &  & 10$^{-3}$ &             & 10$^{14}~$           & L$^{\rm{max}}_{_{\rm{OH}}}$   \\
      &  &           & km s$^{-1}$ &           cm$^{-2}$~ & $h^{-2}_{75}$L$_{\odot}$ \\
\hline
J0034+3025 &  2002 Dec  &2.0  & 13.6 & $<$13.5 & $<$1.99    \\
J0645+2121 &  2002 Dec  &0.79 & 15.3 & $<$5.3  & $<$2.89    \\
            &  2004 Sep  &     &      &         &         \\
J0725$-$0054& 2002 Dec  &2.1  & 12.6 & $<$14.2 & $<$2.48    \\
J0901+2901 &  2002 Nov  &0.46 & 14.1 & $<$3.1  & $<$2.34    \\
            &  2002 Dec  &     &      &         &         \\
J1124+1919 &  2002 Nov  &0.61 & 13.4 & $<$1.1  & $<$1.96    \\
            &  2003 Jan  &     &      &         &         \\
J1347+1217 &  2002 Nov  &0.37 & 12.4 & $<$3.3  & $<$2.25    \\
            &  2003 Jan  &     &      &         &         \\
J1643+1715 &  2004 Mar  &0.44 & 13.3 & $<$3.0  & $<$2.42    \\
J2250+1419 &  2002 Dec  &0.38 & 15.1 & $<$2.6  & $<$2.41    \\
J2316+0405 &  2002 Dec  &0.76 & 14.7 & $<$10.9 & $<$3.03    \\
\hline
\end{tabular}
\end{center}
Col. 1: source name; col. 2: dates of the observations; col. 3 \& 4: 1-$\sigma$
noise for the OH fractional absorption, and the corresponding velocity resolution; cols. 5 \& 6: upper limits on
OH column density and luminosity, respectively.
\label{obsArOH}
\end{table}

%%%%%%%%%%%%%%%%%%%%%%%%%%%%%%%%%%
\section{Summary}
We have presented results from a search for associated H{\sc i} absorptions towards 27 radio sources using
the Arecibo and GMRT telescopes.  We detect H{\sc i} absorption towards six and H{\sc i} emission from one.
A totally new H{\sc i} absorption system was detected against the radio galaxy 3C258, whose HST image shows
an odd-shaped galaxy with an arc-like structure  of high surface brightness, and a larger, fainter tail
extending to the northeast. The H{\sc i} absorption spectrum is complex, possibly due to interaction of the
compact radio source with the ISM of the host galaxy.

We have also presented the results of OH observations towards 9 sources using the Arecibo telescope.
No OH was detected towards any of these sources either in emission or absorption.  The estimated column
densities for H{\sc i} and OH imply an abundance ratio of N(H{\sc i})/N(OH)$\gapp4\times10^{6}$ for 3C258,
adopting typical values for H{\sc i} and OH temperatures and assuming similar covering factors.

We combine our H{\sc i} results with similar, previously published, searches to obtain a `full sample'
of 96 radio sources consisting of 27 GPS, 35 CSS, 13 CFS and 21 LRG radio sources.  The resultant sample has
sources with 5-GHz integrated powers spread over $\sim$6 orders of magnitude and with redshifts up to $z\sim$1.4.
In the `full sample' we find that the H{\sc i} absorption detection rate is highest ($\sim$45\%) for the compact
GPS sources and least ($\sim$25\%) for the large LRG sources.  The anticorrelation between the H{\sc i} column
density and source size for GPS and CSS sources, first noted by Pihlstr\"om et al. (2003), is confirmed by
this enlarged sample. This anticorrelation is also consistent with the optical spectroscopic results of
van Ojik et al. (1997) as described in Section~\ref{NHI-z}.

The H{\sc i} column density, as probed by 21-cm absorption, shows no significant dependence on redshift
for the `full sample' or for the sub-sample of GPS and CSS sources, suggesting that the environments on these
scales are similar for different redshifts.
This is consistent with finding no significant dependence of polarization asymmetry of the lobes of these
sources on redshift (Saikia \& Gupta 2003), and also the fraction of these small sources being
similar at different redshifts (R\"ottgering et al. 1996).

For the subsample of CSS and GPS sources, the H{\sc i} detection rate is higher ($\sim$40\%) for galaxies
than for quasars ($\sim$20\%).  This is consistent with the unification scheme.
The principal (strongest) absorption component detected towards GPS sources appears to be blue-shifted
in 65\% of the cases.  This could be due to the interaction of radio sources with the ambient environment.

%%

%%%%%%%%%%%%%%%%%%%%%%%%%%%%%%%%%%%%%

\section*{Acknowledgments}
We thank Rene Vermeulen, the referee, and Ralph Spencer for many useful suggestions and comments.
We thank the staffs of the Arceibo telescope and GMRT for their assistance during our observations.
The Arecibo Observatory is part of the National Astronomy and Ionosphere Center, which is operated by
Cornell University under a cooperative agreement with the National Science Foundation.
The GMRT is a national facility operated by the National Centre for Radio Astrophysics of the Tata
Institute of Fundamental Research.  We thank  numerous contributers to the GNU/Linux group.
This research has made use of the NASA/IPAC Extragalactic Database (NED) which is operated by the Jet
Propulsion Laboratory, California Institute of Technology, under contract with the National
Aeronautics and Space Administration.
%
%%%%%%%%%%%%%%%%%%%%%%%%%%%%%%%%%%%%%%%
%%%%%%
\end{document}